\documentclass[a4paper,fleqn,usenatbib,useAMS]{mnras}

\usepackage{graphicx}	
\usepackage{multicol}        
\usepackage{longtable}
\usepackage{natbib}
\usepackage{wrapfig}
\usepackage{array}
\usepackage{hyperref}

\usepackage{bm}		
\usepackage{pdflscape}	

\usepackage{enumerate}
\usepackage[shortlabels]{enumitem}
\usepackage{soul}
\usepackage[dvipsnames]{xcolor}
\usepackage{relsize}
\usepackage{amsmath}	
\usepackage{amssymb}	
\usepackage{booktabs}
\usepackage{multirow}
\usepackage{array}

\usepackage[T1]{fontenc}
\usepackage{ae,aecompl}

\usepackage{newtxtext,newtxmath}

\title[MIGHTEE-H{\sc i} galaxies in cosmic filaments]{MIGHTEE-H{\sc i}: Environmental effects of cosmic filaments on extragalactic H{\sc i} detections in the COSMOS and XMM-LSS fields}

\author[Jung et al.]{S. Lyla Jung$^{1}$\thanks{e-mail: \href{mailto:lyla.jung@physics.ox.ac.uk}{lyla.jung@physics.ox.ac.uk}}, Madalina N. Tudorache$^{2}$, Matt J. Jarvis$^{1, 3}$, Michalina Maksymowicz-Maciata$^{4}$, \newauthor Maarten Baes$^{5}$, Marcin Glowacki$^{6, 7}$, Natasha Maddox$^{4}$, Anastasia A. Ponomareva$^{1, 8}$, Kristine Spekkens$^{9}$, \newauthor Tariq Yasin$^{1}$, Bohan Yue$^{1}$
\\
$^{1}$ Astrophysics, Denys Wilkinson Building, Department of Physics, University of Oxford, Keble Road, Oxford OX1 3RH, UK\\
$^{2}$ Institute of Astronomy, University of Cambridge, Madingley Road, Cambridge CB3 0HA, UK\\
$^{3}$ Department of Physics and Astronomy, University of the Western Cape, Robert Sobukwe Road, 7535 Bellville, Cape Town, South Africa\\
$^{4}$ School of Physics, H.H. Wills Physics Laboratory, Tyndall Avenue, University of Bristol, Bristol BS8 1TL, UK\\
$^{5}$ Department of Physics and Astronomy, Universiteit Gent, Proeftuinstraat 86 N3, B-9000 Ghent, Belgium\\
$^{6}$ Institute for Astronomy, University of Edinburgh, Royal Observatory, Edinburgh, EH9 3HJ, UK \\
$^{7}$ Inter-University Institute for Data Intensive Astronomy, Department of Astronomy, University of Cape Town, Cape Town, South Africa\\
$^{8}$ Centre for Astrophysics Research, School of Physics, Astronomy and Mathematics, University of Hertfordshire, College Lane, Hatfield, AL10 9AB, UK \\
$^{9}$ Department of Physics, Engineering Physics and Astronomy, Queen’s University, Kingston, ON K7L 3N6, Canada
}
    
\begin{document}
\maketitle

\begin{abstract}
Cosmic filaments have been recognised as influencing galaxy evolution, but their connection to galactic cold gas reservoirs is elusive, particularly at low redshifts. We investigate the influence of filaments on neutral hydrogen (H{\sc i}) detections in galaxies at $0.02<z<0.09$, using the MeerKAT International GHz Tiered Extragalactic Exploration (MIGHTEE) survey in the COSMOS and XMM-LSS fields. We measure the fraction of optical galaxies from the Dark Energy Spectroscopic Instrument (DESI) Survey detected in H{\sc i} emission as a function of stellar mass, morphology, local density, and the distance to the nearest filament. We find that galaxies closer to filaments have higher stellar masses, higher local densities, and a larger early-type fraction, all of which affect the H{\sc i} detection. After controlling for these dependencies, filaments have measurable effects on H{\sc i} detection. Late-type galaxies in low-local density environments show reduced H{\sc i} detection fractions close to filaments across all stellar mass ranges, suggesting potential gas depletion through ram pressure stripping and cosmic web detachment. Massive early-type galaxies show an enhanced H{\sc i} detection fraction in filaments compared to the voids, suggesting possible gas replenishment through filamentary accretion or gas-rich mergers. Our results suggest that cosmic filaments influence galactic H{\sc i} gas reservoirs through various competing processes, whose relative importance varies with stellar mass and local density. 
\end{abstract}

\section{Introduction}

Matter in the Universe is distributed in a web-like structure (\citealt{Bond_1996}), known as the cosmic web. 
The large-scale structure (LSS) originates from fluctuations in the early Universe and is amplified throughout cosmic evolution as a result of the gravitational collapse of matter from voids into denser regions such as walls, then into filaments and finally into nodes (\citealt{Zeldovich_1970}). 
The current understanding of the LSS is that it is multi-scale and hierarchical (e.g., \citealt{Sheth_2004, Aragon-Calvo_2010, Aragon-Calvo_2024}), spanning a broad range of length, thickness, density, and gas temperature (e.g., \citealt{Galarraga-Espinosa_2020, Zhu_2021, Bahe_2025}). 
The most prominent and dense filaments tend to connect between galaxy clusters, while faint, low-density tendrils are often associated with individual galaxies in voids (\citealt{Alpaslan_2014, Cautun_2014}). 
This highlights the diversity of global environments on large scales beyond the local environment within virialised halos.

Observational studies show that galaxies in dense environments in general have elliptical morphology (\citealt{Oemler_1974, Davis_1976, Dressler_1980, Postman_1984}), are quiescent (\citealt{Balogh_1997, Poggianti_1999, Pimbblet_2002, Lewis_2002, Kauffmann_2004, Baldry_2006}), deficient in cold gas (\citealt{Haynes_1984, Giovanelli_1985}), and have disturbed H{\sc i} morphology (\citealt{Chung_2009, Serra_2012}). 
Much effort in understanding the environmental effects on galaxies has been focused on galaxy clusters, where the influence of environments dominates over that of characteristics intrinsic to individual galaxies (\citealt{Peng_2010}). 
However, massive clusters are rare objects. Most galaxies in the Universe are in less extreme environments than galaxy clusters, such as small groups (\citealt{Yang_2007}), cosmic filaments (\citealt{Tempel_2014}), or both, as groups often reside along filaments (\citealt{Tempel_2014b}). 
There are indications that group environments affect the observed galaxy star-formation properties, particularly for low-mass galaxies (e.g., \citealt{Walker_2010, Cybulski_2014, Barsanti_2018, Cluver_2020}). 
Even the evolution of galaxies currently within massive clusters is linked to low-density environments they inhabited prior to cluster infall (e.g., \citealt{Jung_2018, Kotecha_2022}).

There is growing observational evidence that galaxy properties are related to their cosmic web environments. 
Galaxies close to filaments are more massive than those farther away (\citealt{Poudel_2017, Kraljic_2018, Laigle_2018, Zhang_2025}). 
When it comes to the colour and star formation rate, some studies show that galaxies in filaments tend to be redder in colour and have suppressed star formation activity (\citealt{Alpaslan_2016, Chen_2017, Malavasi_2017, Laigle_2018, Kraljic_2018, Winkel_2021, Zhang_2025, Zarattini_2025}), while others report enhanced star formation activity in some galaxies in filaments (\citealt{Fadda_2008, Porter_2008, Darvish_2014, Kleiner_2017, Vulcani_2019}). 

Similarly, evidence for the influence of filaments on H{\sc i}-detected galaxies has been mixed.  \citet{Kleiner_2017} perform H{\sc i} spectral stacking and find enhanced H{\sc i} mass fractions for massive galaxies ($\log M_{*}/\rm{M_{\odot}}\geq 11$) in filaments, but no statistically significant differences for lower-mass galaxies ($\log M_{*}/\rm{M_{\odot}}<11$). 
In filaments around the Virgo Cluster, both \citet{Lee_2021} and \citet{Yoon_2025} find no clear dependence between the filament proximity and the H{\sc i} mass fraction. In contrast, \citet{CroneOdekon_2018} show that the H{\sc i} galaxies ($8.5<\log M_{*}/\rm{M_{\odot}}<10.5$) in the ALFALFA H{\sc i} survey (\citealt{Giovanelli_2005}) are more H{\sc i}-deficient in filaments.
At higher redshifts ($0.23 < z < 0.49$), \citet{Sinigaglia_2024} report that filament galaxies are H{\sc i}-rich, based on the signal stacking analysis of the MIGHTEE-H{\sc i} Early Science data (\citealt{Maddox_2021}). 
\citet{Luber_2025} also find higher H{\sc i} fractions for the H{\sc i} detected filament galaxies in the COSMOS H{\sc i} Large Extragalactic Survey (CHILES) at $0.09<z<0.48$. 
Further supporting the role of filaments in bringing H{\sc i} to galaxies, studies of H{\sc i} kinematics suggest potential evidence for gas acquisition in filaments via mergers (\citealt{Tudorache2022}) or accretion of filament gas (\citealt{Tudorache_2025}).

Cosmological-volume galaxy formation simulations demonstrate that filamentary gas accretion is a major gas transport channel from intergalactic medium to galaxies, especially in high-redshift and low-mass galaxies (\citealt{Keres_2005, Dekel_2006, vandeVoort_2011}). Such gas inflows are coherent and rich in angular momentum, as they are shaped by the large-scale matter flow from voids into walls and filaments (\citealt{Pichon_2011}). 
At low redshift, megaparsec-scale filaments can inject hydrodynamic instabilities into the circumgalactic medium of galaxies, instead of directly feeding gas to individual galaxies (\citealt{Keres_2009}). 
Yet, given the multi-scale nature of the cosmic web, galaxies in low-density void regions are still likely to be attached to low-density filaments or tendrils even at low redshifts, similar to primordial filamentary networks replenishing gas (\citealt{Aragon-Calvo_2019}).

On the other hand, several mechanisms within filaments are suggested to either directly remove gas from galaxies or gradually cut off the gas supply, which can eventually lead to gas depletion and quenching of star formation. Filaments are filled with mildly shock-heated gas (e.g., \citealt{Eckert_2015, Tanimura_2019}) with a higher density compared to the intergalactic medium in the void regions. Therefore, low-mass galaxies with significant relative velocity with respect to the surrounding large-scale structure can experience ram pressure stripping within filaments (\citealt{Benitez-Llambay_2013, Herzog_2023}). 
The cosmic-web detachment model (\citealt{Aragon-Calvo_2019}) explains the formation of the quenched galaxy population as a result of the hierarchical assembly of cosmic filaments. In this framework, primordial filaments that supply gas to galaxies detach when the system merges with larger cosmic structures, such as other galaxies, halos, and filaments. Further intensifying the situation, gas inflow from the circumgalactic medium to galaxies becomes increasingly inefficient after the filament shell crossing events (\citealt{Song_2021}). 
Galaxies cut off from an additional supply of gas gradually use up the remaining reservoir and become gas-deficient.

In this work, we study how cosmic filaments affect the distribution of H{\sc i}-containing galaxies by measuring the fraction of optical galaxies detected in H{\sc i} emission. 
By controlling for stellar mass, morphology, and local density, we isolate the influence of the global cosmic-web environment, quantified by the distance to the nearest filament, on regulating galaxy gas reservoirs. 

The paper is structured as follows.
We introduce the observational data used in this study in Section \ref{sec:data}. We present our analysis of the H{\sc i} detections depending on the redshift (Section \ref{sec:redshift}), intrinsic galaxy properties (Section \ref{sec:intrinsic}), local density (Section \ref{sec:local_den_effects}), and finally, the cosmic filament environment (Section \ref{sec:fil_effects}). The physical interpretation and implication of the results are presented in Section \ref{sec:discussion}. We summarise our results in Section \ref{sec:summary}. 
We assume a $\Lambda$CDM cosmology based on \citet{Planck_2016} results throughout this work: $\Omega_{\rm m} = 0.309$, $\Omega_{\rm \Lambda} = 0.691$, $\Omega_{\rm b} = 0.0486$, $H_{\rm 0} = 67.8\,\rm km\,s^{-1}Mpc^{-1}$, and $\sigma_{\rm 8} = 0.82$.

\section{Data}\label{sec:data}

\subsection{MIGHTEE-H{\sc i}}

\begin{figure}
    \centering
    \includegraphics[width=\columnwidth]{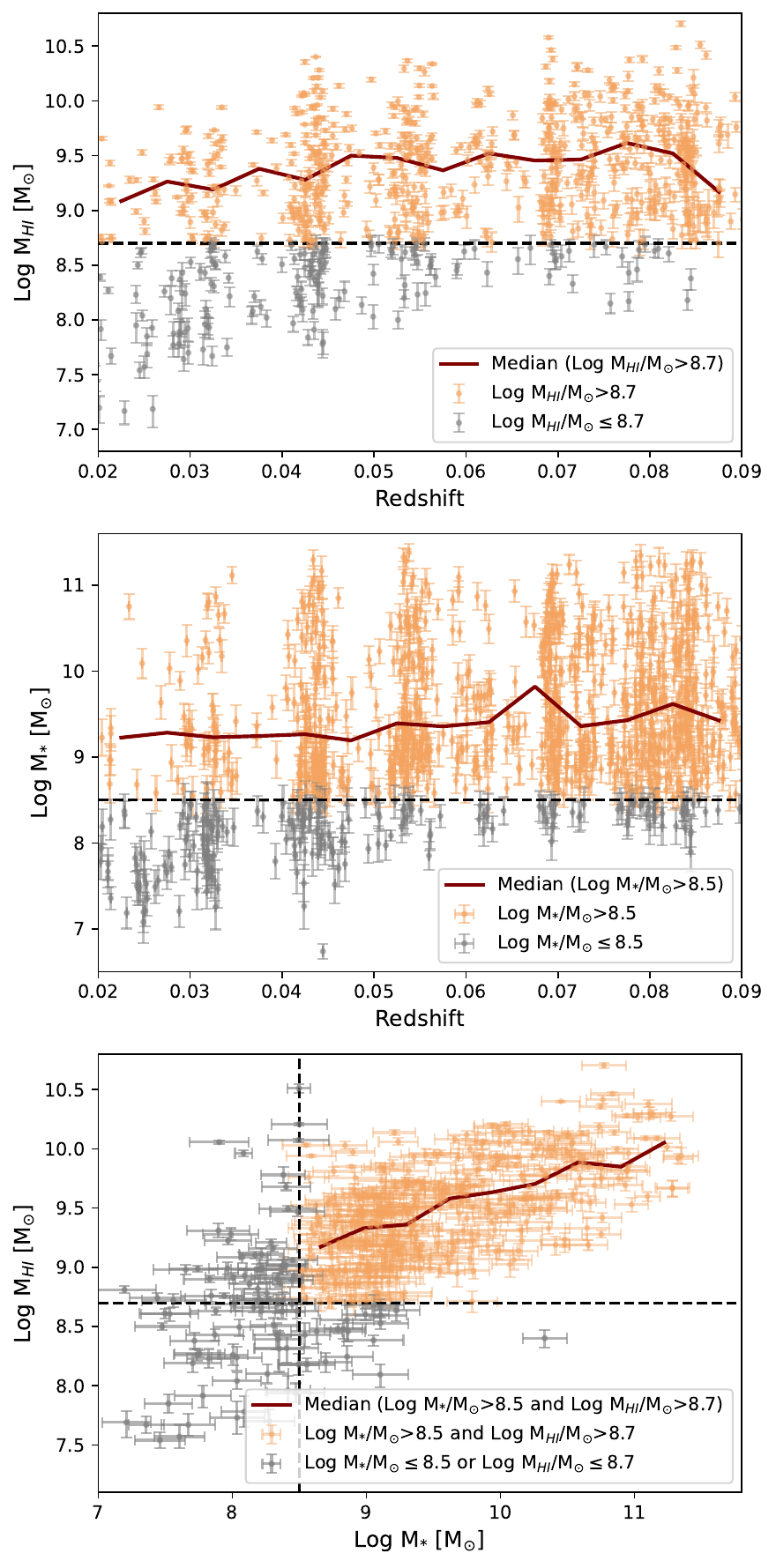}
    \caption{
    Top panel: H{\sc i} mass as a function of redshift. Brown dots are galaxies satisfying the H{\sc i} mass limit adopted in this study ($\log M_{\rm HI}/\rm{M_{\odot}}>8.7$; horizontal dashed line). 
    The solid line shows their median H{\sc i} mass in redshift bins of $\Delta z=0.005$. The grey dots are galaxies below the mass limit.
    Middle panel: Same format as the upper panel, showing the stellar mass of the optical sample as a function of redshift and the stellar mass limit ($\log M_{*}/\rm{M_{\odot}}>8.5$. Bottom panel: stellar mass vs H{\sc i} mass of the optical--H{\sc i} matched sources. 
    }  
    \label{fig:mass_limit}
\end{figure}

The MeerKAT International GHz Tiered Extragalactic Exploration (MIGHTEE; \citealt{Jarvis_2016}) is an extragalactic radio survey program on the South African MeerKAT telescope. 
The survey provides radio continuum (\citealt{Heywood_2022, Hale_2025}), H{\sc i} (\citealt{Maddox_2021, Heywood_2024}), and polarisation (\citealt{Taylor_2024}) data products in four extragalactic fields, the Cosmic Evolution Survey (COSMOS), the XMM-Newton Large-Scale Structure (XMM-LSS), the Extended Chandra Deep Field South (E-CDFS), and the European Large Area ISO Survey South 1 (ELAIS-S1), totalling $\sim 20\,\rm deg^{2}$ sky area. The observations are performed in the wide-band fine mode of MeerKAT, i.e., 32k channels ($26.123\,\rm kHz$ spectral resolution).

This study uses MIGHTEE-H{\sc i} DR1 source catalogues in the COSMOS (\citealt{Maksymowicz-Maciata_2026}) and XMM-LSS fields (Heywood et al. in prep) over the frequency range between $1290-1420\,\rm MHz$. 
The source finding is performed by applying the Line Emission Source-Hunting Integrator\footnote{\url{https://github.com/misia-mm/LESHI}} (LESHI) to the MIGHTEE-H{\sc i} cubes. 
The per-channel noise levels in the centre of the H{\sc i} cubes are $70\,\rm \mu Jy/beam$ in COSMOS and $85 \,\rm \mu Jy/beam$ in XMM-LSS due to a different mosaicking strategy. 
The signal-to-noise threshold used for source identification is 3.5 in the integrated images and 2.5 in the single-channel image. 
Other input parameters for LESHI are presented in \citet{Maksymowicz-Maciata_2026}. 

MIGHTEE's mosaicking strategy leads to nearly uniform sensitivity throughout the fields, except at the edges of the mosaic. 
We limit the analysis to high-sensitivity regions by masking the area within $0.25^{\circ}$ to the edge of the fields.

The top panel of Fig. \ref{fig:mass_limit} shows H{\sc i} mass and redshift of the sources. 
For the analysis presented in this paper, we apply a H{\sc i} mass lower limit of $\log M_{\rm HI}/\rm{M_{\odot}}>8.7$ (dashed horizontal line) to form a sample of H{\sc i} sources with consistent mass limit throughout the redshift range (brown dots with error bars). 
The solid line shows the median $M_{\rm HI}$ of these sources at a redshift interval of 0.005. There is no significant redshift evolution of H{\sc i} mass observed in our sample.
The grey points are H{\sc i} sources below $\log M_{\rm HI}/\rm{M_{\odot}}\leq8.7$. 
The final H{\sc i} sample has 165 and 494 sources in the COSMOS and XMM-LSS fields, respectively.

\subsection{DESI}

\subsubsection{The Legacy Imaging Surveys}

The DESI Legacy Survey (\citealt{Dey_2019}) is an optical photometric survey motivated to provide the target source catalogues for the subsequent spectroscopic survey. 
The survey footprint, totalling $\approx 14,000\,\rm deg^{2}$ sky area, is covered with a compilation of three independent surveys: the Beijing-Arizona Sky Survey (BASS; \citealt{Zou_2017}), the Mayall z-band Legacy Survey (MzLS) and the DECam Legacy Survey (DECaLS). 
The COSMOS and XMM-LSS fields we investigate in this study are located in the DECaLS field.

The optical sources are identified and measured using the {\sc Tractor} algorithm (\citealt{Lang_2016}). 
In this paper, we use the photometric properties for (i) morphological classification of galaxies based on the Sérsic index and (ii) including galaxies with photometric redshifts from the DESI photometric redshift sweep catalogues (\citealt{Zhou_2023}) when measuring the local galaxy number density (see Section \ref{sec:local_density}).

\subsubsection{The stellar mass and emission line catalogue}\label{sec:desi}

One of the key parameters investigated in this study is the three-dimensional distance between galaxies and filaments. Thus, accurate redshift measurements are essential. 
We therefore construct our galaxy sample from the DESI spectroscopic catalogue that provides high-precision spectroscopic redshifts.

The DESI stellar mass and emission line value-added catalogue (\citealt{Zou_2024}) is released as part of DESI DR1 (\citealt{DESI_2026}). 
The catalogue provides emission line and stellar population synthesis model properties for all DESI DR1 spectroscopic sources with the spectral type classified as galaxies and with reliable redshift measurements (`SPECTYPE'==`GALAXY' and `ZWARN'==0).

The spectral energy distribution (SED) of each galaxy is constructed using data from both DESI DR1 spectroscopy and the Legacy Survey photometry. 
For the photometric SED, $g$, $r$, $z$ optical bands and WISE W1 and W2 bands are used. 
In this study, we ensure that the photometric data of the sample galaxies are reliable by using only sources where `RELEASE' > 0 and `SHAPE\_R' > 0. 
Instead of using the full spectroscopic SED for the stellar mass measurements, \citet{Zou_2024} define ten boxcar-shaped broadbands and convolve them with the observed spectra to perform artificial broadband spectrophotometry.
With the five bands and the ten artificial bands across the spectral coverage, they perform the population synthesis fitting using the Code Investigating GALaxy Emission (CIGALE version 2022.1; \citealt{Boquien_2019, Yang_2020, Yang_2022}) assuming the simple stellar population (\citealt{Bruzual_2003}) and \citet{Chabrier_2003} stellar initial mass function. 
The stellar mass and star formation rate are two outputs of CIGALE that will be used extensively throughout this work. 
The median errors of stellar mass and star formation rate of galaxies used in this study are $(\log M_{\rm *})_{\rm err}\approx0.2\,\rm dex$ and $\textit{SFR}_{\rm err}\approx0.3\,\rm M_{\odot}\,yr^{-1}$. 

We use the CIGALE star formation rate to define the star-forming population as galaxies with star formation rates within 0.5 dex of the SDSS star-forming main sequence (\citealt{Zahid_2012}), but the exact definition does not affect the overall results presented in this work. In the final optical sample, there are 460 star-forming galaxies (53\% of the optical sample).

The stellar mass and emission line catalogue includes sources identified in the DESI main survey as well as the survey validation (SV) phases and special observing runs\footnote{\url{https://data.desi.lbl.gov/doc/releases/edr/}}. 
The COSMOS field is covered over four observing phases (main, SV1, SV3, and Special).
The sources from the phase SV3 are those observed in the DESI early data release (the one-per-cent Survey; \citealt{Desi_2024}). 
Different survey phases, especially the survey validation phases, have different observation strategies and target selection criteria (\citealt{Myers_2023}). For example, the SV1 observations are deeper, and the target criteria are broader than other phases, as the purpose of SV1 is to validate the target selection strategy.
For configuring the optical galaxy sample for this study, we consider galaxies from all observing phases to increase the completeness of the sample that match our selection criteria described below. The XMM-LSS field is only covered by the main survey phase.

\subsection{Optical galaxy sample selection}\label{sec:optical_sample}

The key statistic used in this study is the fraction of optical galaxies detected in H{\sc i} emission. 
For this purpose, we define a base sample of optical galaxies from the DESI stellar mass catalogue within the MIGHTEE-H{\sc i} COSMOS and XMM-LSS footprints. 

Duplicated optical sources are removed using the following procedure. (i) If there are multiple sources with identical object ID (`BRICKID' and `BRICK\_OBJID'), only the most massive source is kept in the catalogue.
(ii) Substructures of large galaxies are sometimes identified as separate sources. We remove any sources within the half-light radius of the main component (i.e., the most massive source of the complex) to avoid counting the substructures as separate galaxies.

The middle panel of Fig. \ref{fig:mass_limit} shows the stellar mass of the $r$-band magnitude-limited ($r< 21\,\rm mag$) optical sample as a function of redshift. The excess of galaxies at $z\approx 0.03$, 0.045, 0.055, and 0.07 are overdensities in the LSS within the field. 
In the following analyses, we apply the stellar mass limit of $\log M_{*}/\rm{M_{\odot}}>8.5$ (dashed line) to ensure a stellar-mass-complete galaxy sample (brown dots with error bars) over the redshift range of interest. The solid line shows the median stellar mass of these galaxies in redshift bins with an interval of 0.005. No significant redshift evolution of stellar mass is observed for the mass-limited sample. 
Galaxies below $\log M_{*}/\rm{M_{\odot}}\leq8.5$ are shown in grey dots. 

For the 865 optical galaxies selected as a final sample, we perform matching with the H{\sc i} catalogues to evaluate their H{\sc i} detection. 
Each H{\sc i} source is matched to an optical galaxy if the angular separation is smaller than the extent of the H{\sc i} source\footnote{We measure the largest distance between two points belonging to the $3\sigma$ contour of the H{\sc i} emission.} and the redshift difference is smaller than $\Delta z<0.001$. 
If there is more than one galaxy in the distance and redshift range, the one with the smallest angular separation is matched, unless the H{\sc i} source is visually classified as a blend of H{\sc i} emission from multiple galaxies. In the latter case, all the optical galaxies within the spatial extent and the redshift range are assigned to the H{\sc i} source. We present the stellar mass--H{\sc i} relation of the optical--H{\sc i} matched sources in the bottom panel of Fig. \ref{fig:mass_limit}. The two mass limits, $\log M_{\rm HI}/\rm{M_{\odot}}>8.7$ and 
$\log M_{*}/\rm{M_{\odot}}>8.5$, are shown as horizontal and vertical dashed lines, respectively. We show sources satisfying both limits with brown dots and error bars. The solid line shows the median $\log M_{\rm HI}$ at a given $\log M_{*}$ of these sources. 

Note that not all H{\sc i} sources in the  H{\sc i} catalogue are matched with optical galaxies. Of the 165 (494) H{\sc i} sources identified in the COSMOS (XMM-LSS) field, 149 (294) are matched with an optical spectroscopic source. The matched fraction is higher in the COSMOS field (90\%) than in the XMM-LSS field (60\%), at least in part due to the high observation completeness of DESI in the COSMOS field. For reference, the DESI SV3 observation tile covers the entire COSMOS field in one pointing; thus, there is no loss of galaxies due to a gap in the DESI observation tiling. This is not the case in the XMM-LSS field, which covers a larger area. 
Nevertheless, assuming there is no significant bias in the galaxy population due to the incomplete sampling, this should not affect the H{\sc i} detection fraction analysis we present in the later part of this paper. 
We confirm that H{\sc i} sources unmatched with optical galaxies often have a visually identifiable optical counterpart in the Legacy Surveys multi-band images, which did not meet our optical sample selection criteria, e.g., below the stellar mass limit, or not in the DESI DR1 coverage as explained above. These galaxies are not included in the analysis.

\subsection{Cosmic filament catalogue}\label{sec:filament_finding}

\begin{figure*}
    \centering
    \includegraphics[width=0.7\textwidth]{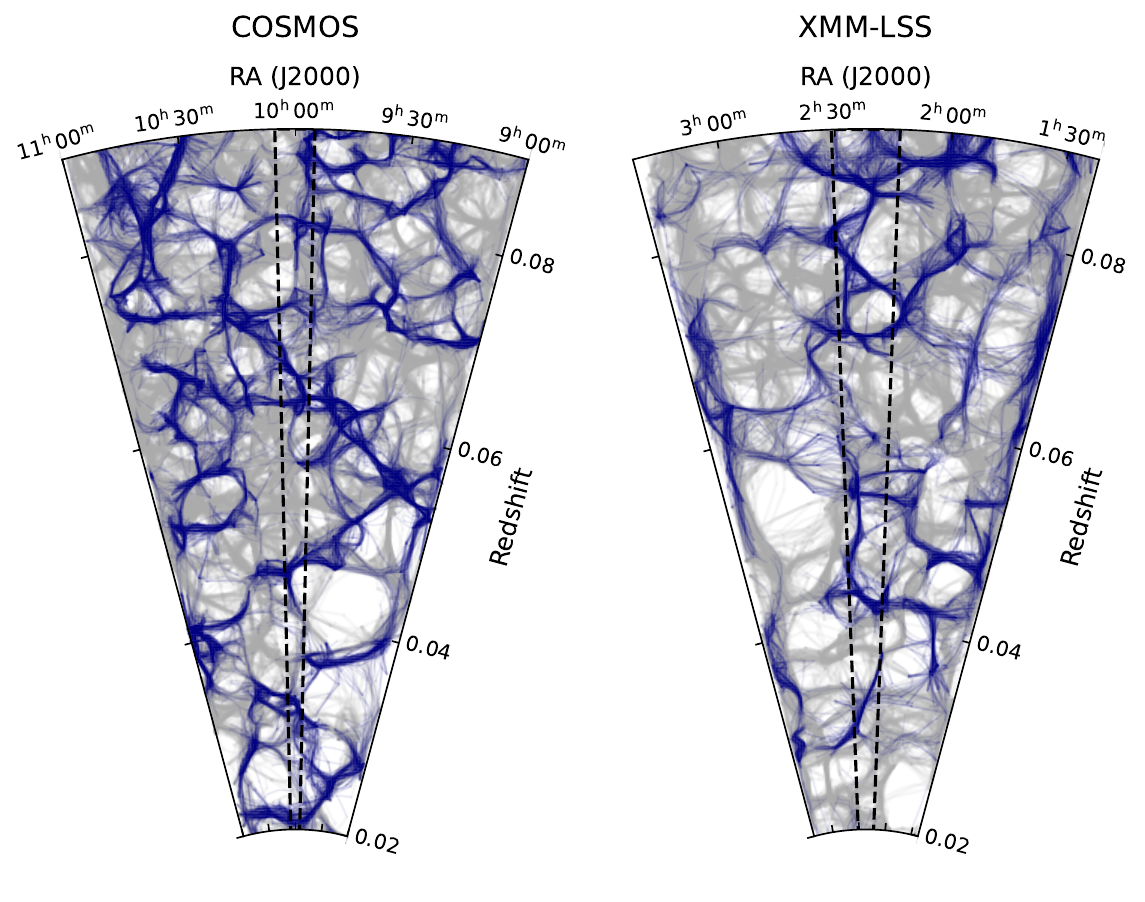}
    \caption{
    Distribution of cosmic filaments identified with DisPerSE in the COSMOS (left) and XMM-LSS (right) fields. The blue lines show filaments within $\pm1^{\circ}$ of the declination centre of each MIGHTEE-H{\sc i} field. The low opacity of the lines is set to highlight regions where filaments from the 100 jackknife realisations overlap. The grey lines show the remaining filaments in the $30^{\circ}\times30^{\circ}$ fields. The vertical dashed lines are the RA limits of the MIGHTEE-H{\sc i} fields. 
    }  
    \label{fig:filaments}
\end{figure*}

As the underlying dark matter distribution of the LSS is not directly observable, we use the galaxy distribution to identify the cosmic web. 
We apply the Discrete Persistent Structure Extractor (DisPerSE; \citealt{Sousbie_2011, Sousbie_2011b}) to the 3D distribution of DESI spectroscopic galaxies within a wide area ($30^{\circ}\times30^{\circ}$) centred on each MIGHTEE-H{\sc i} field up to redshift 0.1. This minimises unrealistic boundary effects when assigning higher-redshift ($z\sim 0.09$) galaxies or galaxies close to the field edges to filaments. 
Because we construct the filament catalogue over a wider redshift range compared to the redshift range of the science sample introduced in Section \ref{sec:optical_sample}, we use a higher stellar mass limit for filament identification ($\log M_{*}/\rm{M_{\odot}}>9$). 
The size of the field used for filament finding does not affect the results of this work, as long as it is sufficiently larger than the size of the MIGHTEE-H{\sc i} footprints. 

For the input galaxy catalogues, we prioritise homogeneous galaxy sampling over maximising completeness. Regions observed multiple times over different DESI observation phases can result in an unrealistic concentration of galaxies, which affects the filament finding.
Thus, we configure the input catalogue only using the main survey and the SV3 phase for the COSMOS field, since their observing specifications are nearly the same (\citealt{Myers_2023, DESI_2024b}). Duplicate sources between the main survey and SV3 are removed from the input. In the XMM-LSS field, we use the main survey available.

Note that we only use the spectroscopic sample of galaxies for the filament identification. Therefore, the redshift uncertainty of the input galaxies is negligible.

Before applying DisPerSE to the input galaxy catalogue, we modify the redshift of galaxies in large groups to mitigate the redshift-space distortion due to the peculiar velocities (i.e., the Fingers-of-God effect; \citealt{Jackson_1972}) as follows. 
For any groups in the DESI Halo-based Group Catalogue (\citealt{Yang_2021}) with a halo mass larger than $10^{12}\,\rm M_{\odot}$ and richness larger than 5, we identify member galaxies within $\pm \Delta z c =  3000\,\rm km\,s^{-1}$. The angular extent of each halo is defined as the 95th percentile of the angular separations to all associated galaxies in the Galaxy-group Relationship Catalogue provided. 
Then, following \citet{Kraljic_2018}, we compress the redshift of the member galaxies so that the radial dispersion is equal to the transverse dispersion of the group in the physical scale.



DisPerSE requires several input parameters that decide the overall shape and length of the filaments in the output. 
We choose the mirror boundary condition, although this does not affect the results presented in this paper, since our field of interest is much smaller than the input catalogue for filament identification. 
We adopt the persistence threshold of $3.5\sigma$ motivated by the discussion in \citet{Sousbie_2011b} that a threshold above $>3\sigma$ is generally enough to remove spurious filaments of small significance. 
We perform three rounds of skeleton smoothing to remove any unphysical sharp bending of the identified filaments.

Following \citet{Tudorache_2025}, we construct 100 input galaxy catalogues jack-knifed from the full sample, each randomly omitting 5 per cent of the sample. 
Fig. \ref{fig:filaments} illustrates the 100 filament realisations in the RA--redshift space centred at the COSMOS (left) and XMM-LSS (right) fields. 
The blue lines are filaments within a slice through the declination centre of each field, with $\pm1^{\circ}$ thickness. The opacity of the lines is set low to highlight the regions where filaments from different jack-knife realisations overlap. The grey lines in the background are the remaining filaments outside the slice. The dashed lines along the redshift axis show the RA limits of the MIGHTEE-H{\sc i} fields, for reference.

The output of DisPerSE describes each filament as a sequence of sampling points. 
We use the coordinates of these sampling points to identify the nearest filament from each galaxy and measure the 3D distance between the galaxy and the filament ($D_{\rm fil}$). In addition to filaments, DisPerSE provides the coordinates of critical points, i.e., where the gradient of the Delaunay Tessellation density field is zero. We define the maxima and bifurcation points as nodes of the filamentary network and measure the 3D distance to the nearest node of each galaxy ($D_{\rm node}$). 
From the 100 jack-knifed filament catalogues, we get 100 measurements of $D_{\rm fil}$ and $D_{\rm node}$ for each galaxy. To assign representative $D_{\rm fil}$ and $D_{\rm node}$ values, we estimate the probability distribution of the parameters using a Gaussian kernel density estimator (KDE) and take the most probable value across all realisations. 
The uncertainties associated with each representative $D_{\rm fil}$ and $D_{\rm node}$ are typically $\approx 1.3$ and  $2.2\,\rm Mpc$, respectively, characterised by half the difference between the 16th and 84th percentiles of the probability distribution. 
In the following sections, we adopt $D_{\rm fil}$ and $D_{\rm node}$ as quantitative measures of the cosmic web environment of galaxies. 

\begin{figure}
    \centering
    \includegraphics[width=\columnwidth]{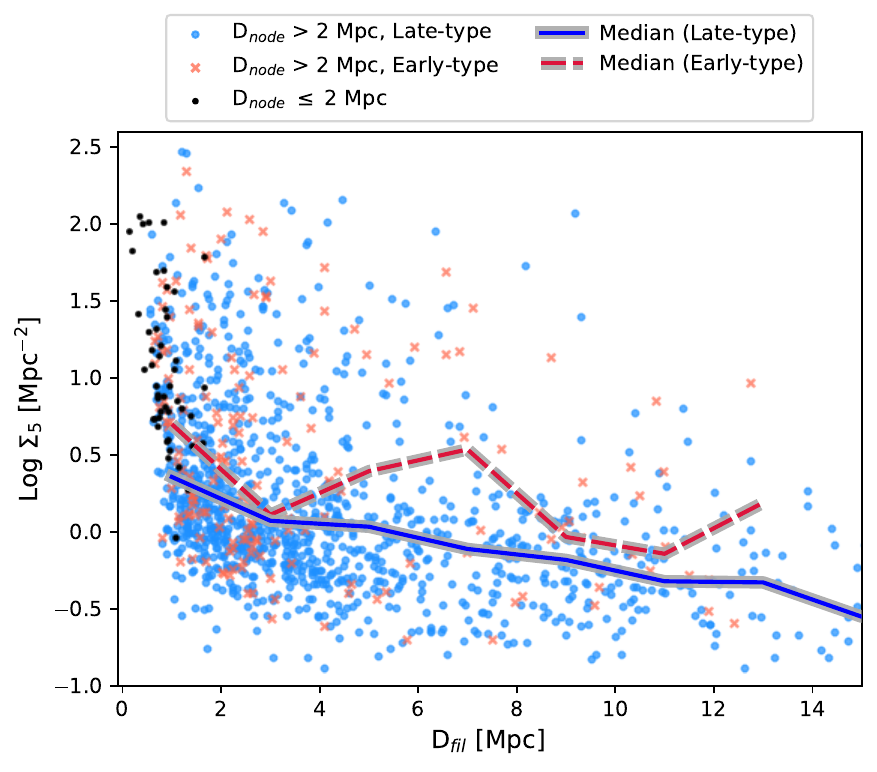}
    \caption{
    The local density ($\log \Sigma_{5}$) and the distance to the nearest cosmic filament ($D_{\rm fil}$) of the late-type (blue dot) and early-type (red cross) galaxies. 
    The solid and dashed lines show the median $\log\Sigma_{5}$ of late- and early-type galaxies as a function of $D_{\rm fil}$. The black dots are galaxies within $D_{\rm node}\leq2\,\rm Mpc$.}  
    \label{fig:envs}
\end{figure}

\subsection{Local density measurement}\label{sec:local_density}

We use the projected galaxy number density ($\Sigma_{N}$) based on the distance to the fifth nearest neighbour as a tracer of the local environment:
\begin{equation}
    \Sigma_{N} = \frac{N}{\pi d_{N}^{2}},
\end{equation}
where $N=5$ and $d_{N}$ is the projected distance to the $N$th nearest neighbour galaxy within the redshift range of $\Delta zc=\pm1000\,\rm km\,s^{-1}$ (\citealt{Baldry_2006,Brough_2013}). 
Neighbouring galaxies are determined using the spectroscopic galaxy sample before the stellar mass cut explained in Section \ref{sec:optical_sample}, along with the photometric redshift sample from the DESI sweep catalogues (\citealt{Zhou_2023}; hereafter photo-z catalogues). 
As the photometric redshift catalogues include both stars and galaxies, we adopt the star-galaxy separation criteria used for the DESI Bright Galaxy Survey (\citealt{Hahn_2023, DESI_2024b}); any object that is not in the \textit{Gaia} DR2 catalogue (\citealt{Gaia_2018}) or that satisfies $G_{\rm Gaia}-r > 0.6$, where $G_{\rm Gaia}$ is the \textit{Gaia} $G$-band magnitude and $r$ is the raw $r$-band magnitude before the galactic extinction correction, is considered a galaxy and included in our local density estimation. 
Duplicated galaxies in both the spectroscopic and photometric catalogues are identified using the DESI object ID and removed from the photometric catalogue. 
We apply an upper limit of the $r$-band absolute magnitude $M_{r}<-17.5\,\rm mag$ for this procedure so that the luminosity limit of the galaxies used for the local density measurements remains constant over the redshift range considered. 
We calculate the absolute magnitudes using their observed flux and the Galactic extinction coefficients (\citealt{Schlegel_1998}) provided in the DESI catalogues. $K$-corrections are made using the SDSS $K$-corrections calculator (\citealt{Chilingarian_2010, Chilingarian_2012}), following \citet{Ebrova_2025}.

The DESI photo-z catalogue provides redshift measurements from other spectroscopic surveys (`Z\_SPEC'), such as the Sloan Digital Sky Survey (SDSS; \citealt{Strauss_2002}) and the Galaxy And Mass Assembly (GAMA; \citealt{Baldry_2018}), if available. 
For galaxies with existing spectroscopic redshifts, we use those measurements instead of the photometric redshifts. 
For galaxies with only photometric redshift measurements, we construct the probability distribution of the redshift based on the mean and standard deviation (`Z\_PHOT\_MEAN' and `Z\_PHOT\_STD', respectively), assuming the normal distribution and scale it by the cosmic volume at corresponding redshifts. 
Then, we randomly sample the photometric redshift of each galaxy 1000 times, combine them with the spectroscopic catalogue, and calculate $\Sigma_{5}$ for each realisation. The median of the 1000 $\Sigma_{5}$ measurements is used as the representative local density estimate of a galaxy. 

In Fig. \ref{fig:envs}, we show the distribution of $\log\Sigma_{5}$ as a function of $D_{\rm fil}$ for the optical sample used in this study. Galaxies at $D_{\rm node}>2\,\rm Mpc$ are coloured by their morphology (blue dots: late-type; red cross: early-type). 
The solid and dashed lines show the median $\log\Sigma_{5}$ of late- and early-type galaxies in a given $D_{\rm fil}$ range. 
There is an overall anti-correlation between the two environmental measures, such that the local density is higher closer to filaments. Still, we find a wide scatter in the local density inside the filaments, spanning over two orders of magnitude. Early-type galaxies have overall higher local densities than late-type galaxies at a fixed $D_{\rm fil}$. 
Galaxies located close to nodes ($D_{\rm node}\leq2\,\rm Mpc$, black dots) occupy higher local densities than rest of the sample.

\section{Results}\label{sec:HI-detection}

\begin{figure}
    \centering
    \includegraphics[width=\columnwidth]{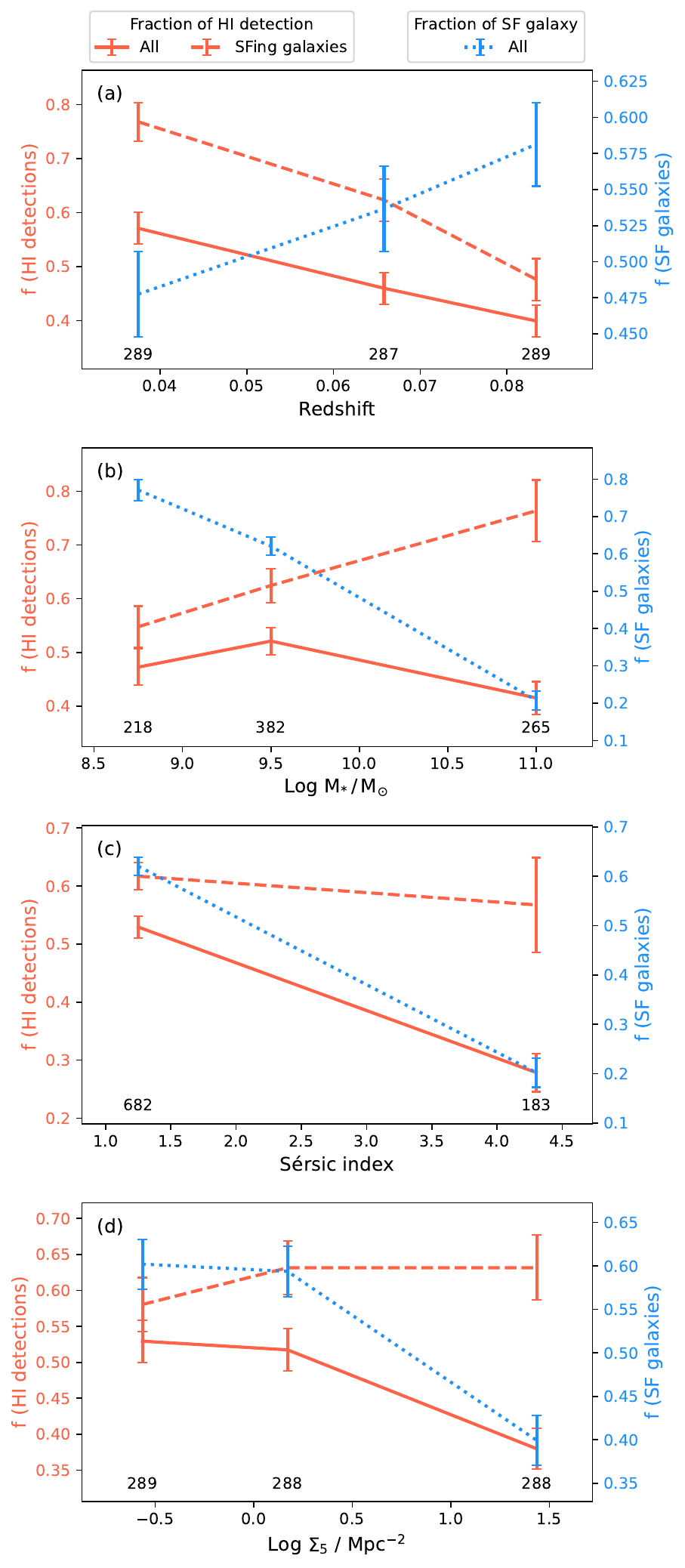}
    \caption{
    H{\sc i} detection fraction (red solid line for all optical galaxies; red dashed line for star-forming galaxies only) and star-forming galaxy fraction (blue dotted line) of optical galaxies as a function of redshift, stellar mass, Sérsic index, and local density (from top to bottom). The error bars show the standard error. The numbers at the bottom of each panel show the total number of optical galaxies in each bin.}  
    \label{fig:mass_detection}
\end{figure}

In this study, we quantify the distribution of H{\sc i} galaxies by measuring the fraction of H{\sc i} detections among optically selected galaxies. 
We first explore how the observed H{\sc i} detection fraction varies with redshift (Section \ref{sec:redshift}), intrinsic galaxy properties, such as stellar mass and morphology (Section \ref{sec:intrinsic}), and the local galaxy number density (Section \ref{sec:local_den_effects}). We then investigate the influence of cosmic filament environments on the H{\sc i} detection while controlling for these parameters.

To first order, the detectability of H{\sc i} in a galaxy depends on the intensity of the H{\sc i} emission line. 
\citet{Maksymowicz-Maciata_2026} present a detailed analysis of how H{\sc i} mass, inclination, and distance to a source affect the completeness of the H{\sc i} source finding, along with a functional description of the completeness fraction at the noise level of the MIGHTEE-H{\sc i} data (see their equation 1). 
In the redshift range of interest in this work ($0.02<z<0.09$), the predicted source finding completeness assuming the noise level of the COSMOS field ($70\,\rm \mu Jy/beam$) is, on average, $95.5\%$ for H{\sc i} sources more massive than $\log M_{\rm HI}/\rm{M_{\odot}} >8.7$, i.e. the H{\sc i} mass limit used in this study. Note that the observed noise level in the XMM-LSS field ($85\,\rm \mu Jy/beam$) is slightly higher than that of the COSMOS field.

\subsection{Redshift effects}\label{sec:redshift}

In panel (a) of Fig. \ref{fig:mass_detection}, we show the H{\sc i} detection fraction ($y$-axis on the left) of all optical galaxies and star-forming galaxies (as defined in Section \ref{sec:desi}) with solid and dashed lines with error bars, respectively, as a function of redshift. The numbers at the bottom of the panel are the number of optical galaxies in each bin. 
The fraction of star-forming galaxies in our sample  ($y$-axis on the right, blue dotted line with error bars) increases mildly with redshift. 
In contrast, the H{\sc i} detection fraction rapidly decreases with increasing redshift, regardless of the star-formation status. This is a direct result of Malmquist bias of the H{\sc i} flux with increasing distance to the sources. The H{\sc i} detection fractions of star-forming galaxies are always higher than the total sample, as expected. 

As we examine how H{\sc i} detection depends on various galactic and environmental properties in the following sections, we ensure that the samples being compared have similar redshift distributions. This alleviates the possibility that differences in the H{\sc i} detection fractions between two samples arise from the redshift effects.

\subsection{Stellar mass and morphology effects} \label{sec:intrinsic}

In panel (b) of Fig. \ref{fig:mass_detection}, we show the star-forming galaxy fraction and the H{\sc i} detection fraction in three stellar mass bins:
$8.5<\log M_{*}/\rm{M_{\odot}}<9$, $9\leq\log M_{*}/\rm{M_{\odot}}<10$, and $10\leq\log M_{*}/\rm{M_{\odot}}$. 
The fraction of star-forming galaxies in our sample decreases significantly with increasing stellar mass (blue dotted line with error bars), from $77\%$ in the lowest mass bin to $20\%$ in the most massive bin. 

When considering star-forming galaxies exclusively (red dashed line), we find that the H{\sc i} detection fraction increases with increasing stellar mass. This trend arises because the H{\sc i} mass of H{\sc i}-detected galaxies is overall larger in more massive galaxies (see the bottom panel of Fig. \ref{fig:mass_limit}). Therefore, massive H{\sc i} galaxies are more likely to be detected at a fixed observation sensitivity. 
This positive scaling relation between H{\sc i} mass and stellar mass has been observed in previous H{\sc i} studies as well (e.g., \citealt{Huang_2012, Maddox_2015, Parkash_2018, Saintonge_2022,   Pan_2023}). 
The H{\sc i} detection fraction of the total sample (red solid line), regardless of their star formation status, reflects a combination of the two trends. 
The intermediate stellar mass bin has the highest H{\sc i} detection fraction. 
The fraction is low in the highest stellar mass bin, as the population is dominated by quiescent H{\sc i}-deficient galaxies. 

Panel (c) of Fig. \ref{fig:mass_detection} shows the H{\sc i} detection fraction and the star-forming galaxy fraction of galaxies with Sérsic index $\leq2.5$ and $>2.5$.
In galaxies with high Sérsic indices, both the H{\sc i} detection and star-forming galaxy fractions are lower than in the low Sérsic index population. 
However, when considering the H{\sc i} detection of star-forming galaxies exclusively (red dashed line), the H{\sc i} detection fraction does not depend on the Sérsic index. 
This is because there is a non-negligible number of gas-rich star-forming galaxies with higher Sérsic indices. 
Our visual investigation suggests that many of these high-Sérsic index \textit{and} star-forming galaxies have late-type morphology. 
It has been suggested by previous studies (e.g., \citealt{Lange_2015}) 
that dividing early- and late-type galaxies based on both morphological and star formation tracers is superior to using only one of the tracers, especially when low-mass galaxies are considered ($\log M_{*}/\rm{M_{\odot}}\lesssim 10$). 
Motivated by this, we define the early-type sample as galaxies that satisfy the Sérsic index $>2.5$ \textit{and} are quiescent in star formation, and the remaining galaxies as the late-type sample. 
For reference, the H{\sc i} detection fractions of early- and late-type galaxies are $0.21\pm0.03$ and $0.53\pm0.02$, respectively.

\subsection{Local density effects}\label{sec:local_den_effects}

\begin{figure}
    \centering
    \includegraphics[width=0.9\columnwidth]{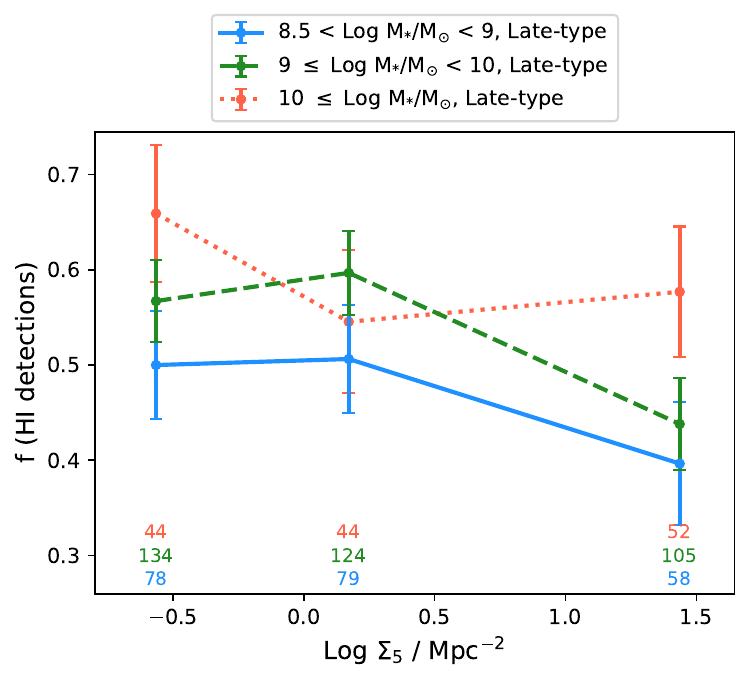}
    \caption{
    H{\sc i} detection fraction of late-type galaxies as a function of local density. Each colour represents different stellar mass bin: $8.5<\log M_{*}/\rm{M_{\odot}} <9$ (blue solid line), $9\leq\log M_{*}/\rm{M_{\odot}} <10$ (green dashed line), and $\log M_{*}/\rm{M_{\odot}} \geq10$ (red dotted line). The error bars show the standard error. The numbers at the bottom show the total number of optical galaxies in each bin.
    }  
    \label{fig:detection_localden}
\end{figure}

The bottom panel of Fig. \ref{fig:mass_detection} shows the H{\sc i} detection and star-forming galaxy fractions depending on the local density ($\log \Sigma_{5}$). 
There is a significant decrease in the fraction of star-forming and H{\sc i}-detected galaxies with increasing local density, in line with the well-known morphology-density relation (\citealt{Dressler_1980}). 
This trend disappears when we limit the sample to star-forming galaxies (red dashed line). 

We further investigate the dependence of H{\sc i} detection on local density by dividing the sample into stellar mass bins. 
Fig. \ref{fig:detection_localden} shows the H{\sc i} detection fraction of late-type galaxies in three stellar mass ranges (blue solid line: $8.5<\log M_{*}/\rm{M_{\odot}}<9$, green dashed line: $9\leq \log M_{*}/\rm{M_{\odot}}<10$, and red dotted line: $\log M_{*}/\rm{M_{\odot}}\geq10$). 
By limiting the sample to late-type galaxies, we test the influence of local density on H{\sc i} detection without morphological transformation. 
For the two lower-mass bins, the H{\sc i} detection fraction decreases at local densities greater than ($\log \Sigma_{5}/\rm{Mpc^{-2}}>0.4$). 
In contrast, galaxies in the highest stellar mass bin show no dependence of the H{\sc i} detection fraction on local density. 
This suggests that the physical mechanisms for removing H{\sc i} gas in galaxies in high-density environments primarily affect low-mass galaxies. 

\subsection{Cosmic filament effects}\label{sec:fil_effects}

In this section, we explore how the H{\sc i} detection fraction varies with the distance to the nearest cosmic filament ($D_{\rm fil}$). 
To separate filament-specific effects from those of massive clusters and groups, we exclude galaxies close to nodes ($D_{\rm node}\leq2\,\rm Mpc$) from the following analysis\footnote{We have tested different distance limits ($D_{\rm node}>1,\,3,$ and $5\,\rm Mpc$) and confirm that the exact choice of the $D_{\rm node}$ limit only marginally changes the results presented in this section and the conclusions remain unchanged. }.

\subsubsection{Filament effects on stellar mass and morphology}

\begin{figure}
    \centering
    \includegraphics[width=\columnwidth]{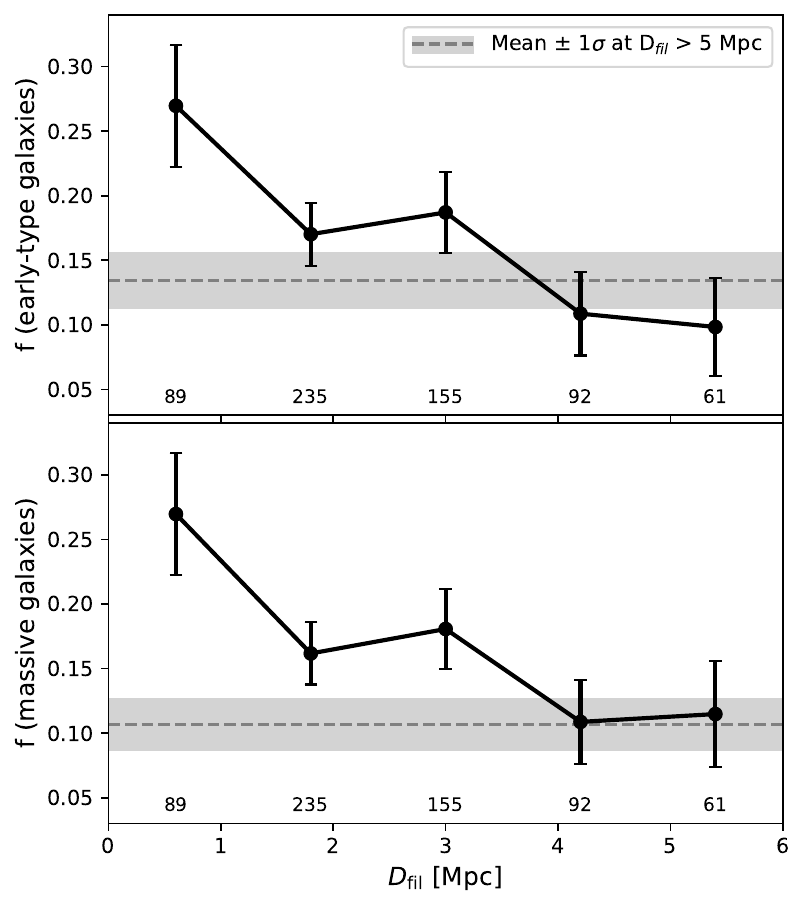}
    \caption{
    Fraction of early-type galaxies (upper panel) and massive galaxies ($\log M_{*}/\rm{M_{\odot}} \geq10.5$; lower panel) at a given distance to the nearest filament, $D_{\rm fil}$. The error bars show the standard error. 
    The dashed line and grey band are the mean and standard error of the reference sample with $D_{\rm fil}>5\,\rm Mpc$. We show the total number of galaxies in each bin at the bottom of the panels. 
    }  
    \label{fig:early_frac}
\end{figure}

\begin{table}
\caption{P-values from the Monte Carlo null hypothesis tests quantifying the significance of the deviation between the observed and reference fractions of early-type galaxies (top) and massive galaxies (bottom) as a function of $D_{\rm fil}$, for individual bins (left) and cumulative $D_{\rm fil}$ ranges (right).}\label{tab:early_p_val}
\centering
\begin{tabular}{cccc}
\hline
\multicolumn{4}{c}{\textbf{f (early-type galaxies) vs $D_{\rm fil}$}} \\ \hline
\multicolumn{2}{c}{Interval statistics} & \multicolumn{2}{c}{Cumulative statistics} \\
$D_{\rm fil}$ {[}Mpc{]} & p-value & $D_{\rm fil}$ {[}Mpc{]} & p-value \\ \hline
{[}0.0, 1.2) & 0.0027 & - & - \\ \hline
{[}1.2, 2.4) & 0.1388 & {[}0.0, 2.4) & 0.0193 \\ \hline
{[}2.4, 3.6) & 0.0830 & {[}0.0, 3.6) & 0.0162 \\ \hline
{[}3.6, 4.8) & 0.7664 & {[}0.0, 4.8) & 0.0431 \\ \hline
{[}4.8, 6.0) & 0.8155 & {[}0.0, 6.0) & 0.0735 \\ \hline\hline
\multicolumn{4}{c}{\textbf{f (massive galaxies) vs $D_{fil}$}} \\ \hline
\multicolumn{2}{c}{Interval statistics} & \multicolumn{2}{c}{Cumulative statistics} \\ 
$D_{\rm fil}$ {[}Mpc{]} & p-value & $D_{\rm fil}$ {[}Mpc{]} & p-value \\ \hline
{[}0.0, 1.2) & 0.0002 & - & - \\ \hline
{[}1.2, 2.4) & 0.0356 & {[}0.0, 2.4) & 0.0015 \\ \hline
{[}2.4, 3.6) & 0.0177 & {[}0.0, 3.6) & 0.0008 \\ \hline
{[}3.6, 4.8) & 0.5132 & {[}0.0, 4.8) & 0.0030 \\ \hline
{[}4.8, 6.0) & 0.4734 & {[}0.0, 6.0) & 0.0050 \\ \hline
\end{tabular}%
\end{table}

\begin{figure*}
    \centering
    \includegraphics[width=\textwidth]{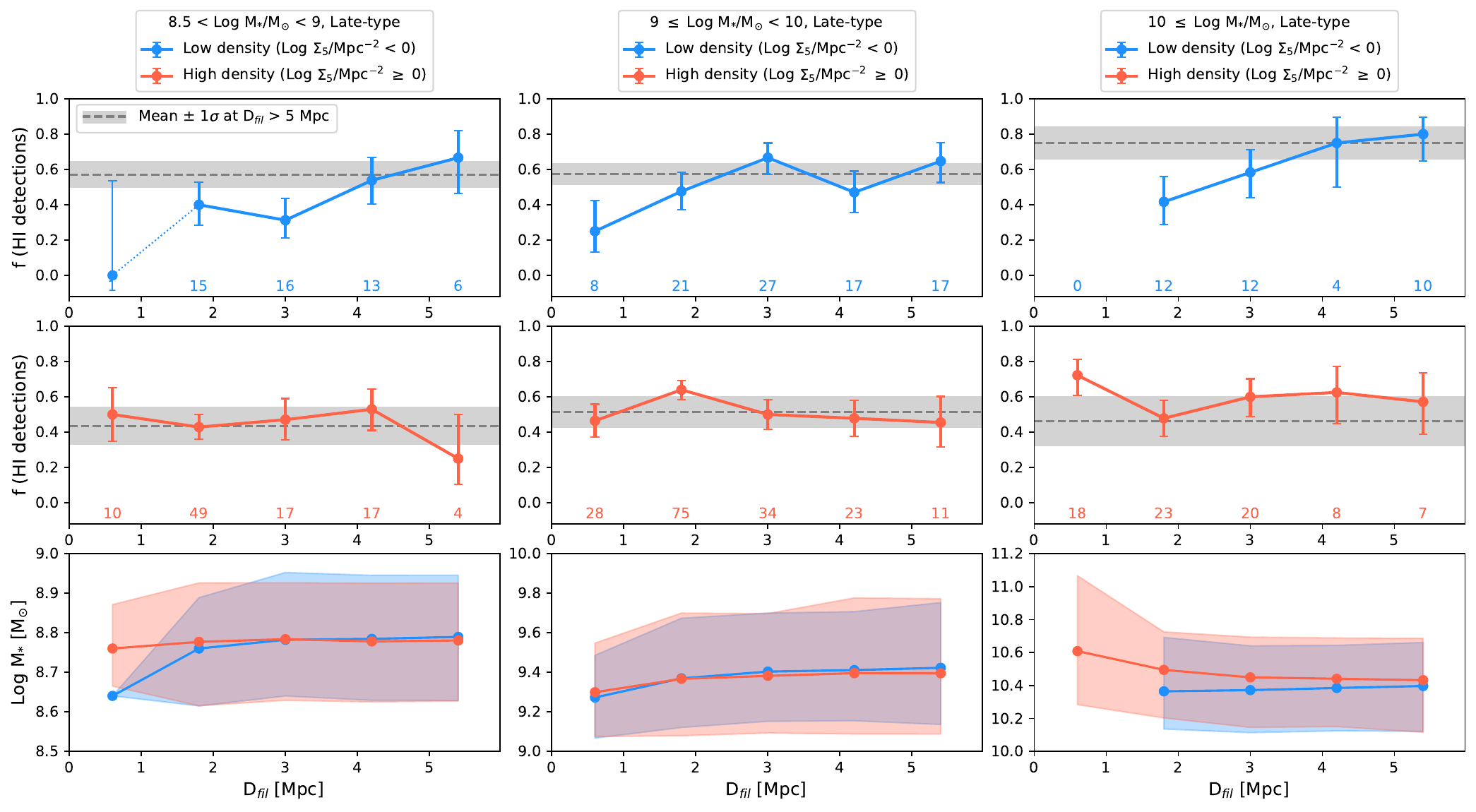}
    \caption{
    Top panels: H{\sc i} detection fraction as a function of $D_{\rm fil}$, for late-type galaxies in different local densities ($\log\Sigma_{5}/\rm{Mpc^{-2}}<0$). The error bars show the 68\% Jeffrey's credible interval.
    The three columns correspond to three stellar mass bins: $8.5<\log M_{*}/\rm{M_{\odot}} <9$ (left), $9\leq\log M_{*}/\rm{M_{\odot}} <10$ (middle), and $\log M_{*}/\rm{M_{\odot}} \geq10$ (right). 
    The numbers at the bottom of the panels indicate the number of galaxies in each $D_{\rm fil}$ bin.
    The horizontal dashed line and band are the H{\sc i} detection fraction of the reference sample at $D_{\rm fil}>5\,\rm Mpc$ and its standard error.
    Middle panels: Same format as the top panels showing the high-local density samples ($\log\Sigma_{5}/\rm{Mpc^{-2}}\geq0$).
    Bottom panels: the mean and $1\sigma$ scatter of the stellar mass as a function of $D_{\rm fil}$, for galaxies used to calculate the H{\sc i} detection fraction in the upper panels. 
    }  
    \label{fig:hi_frac_latetype}
\end{figure*}

\begin{figure}
    \centering
    \includegraphics[width=0.38\textwidth]{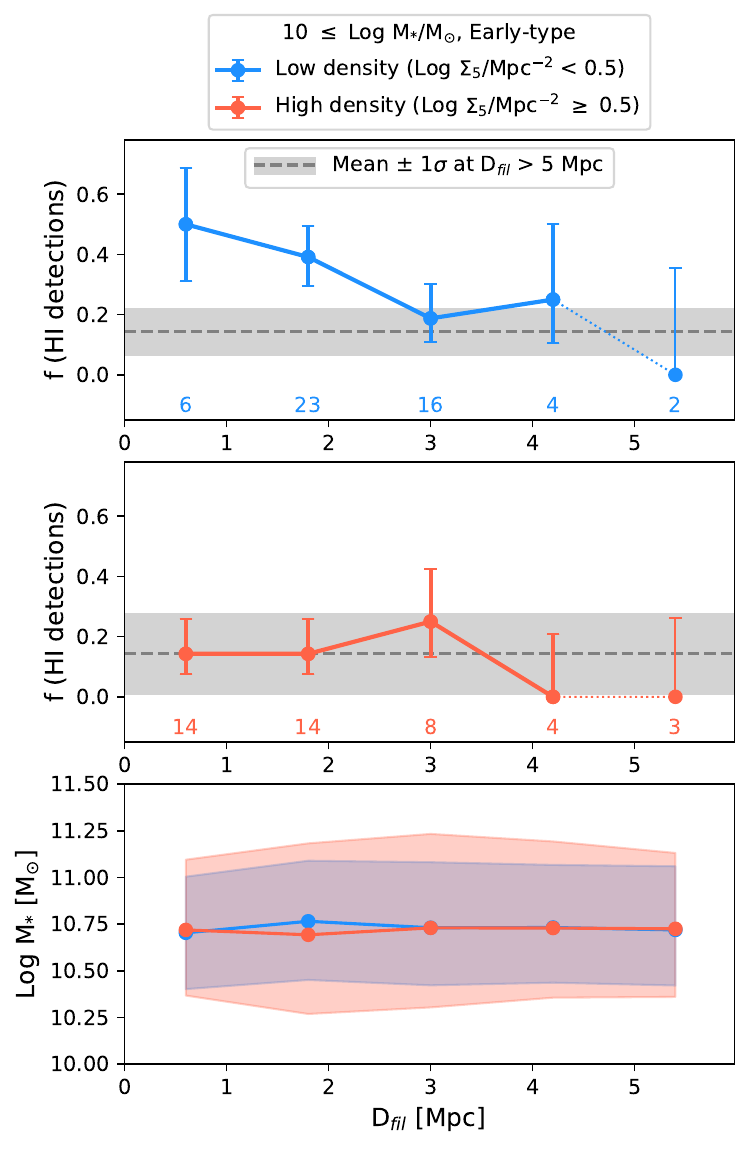}
    \caption{
    The same format as Figure \ref{fig:hi_frac_latetype}, presenting only early-type galaxies at $\log M_{*}/\rm{M_{\odot}}\geq10$. 
    }  
    \label{fig:hi_frac_earlytype}
\end{figure}

Previous studies report that stellar mass, morphology, and star-formation properties of galaxies are related to their cosmic web environments (e.g., \citealt{Alpaslan_2016, Kuutma_2017, Kraljic_2018, Hoosain_2024}).  
We confirm that this is indeed the case in our sample. 
In Fig. \ref{fig:early_frac}, we show the fraction of early-type galaxies (top panel) and massive galaxies above $\log M_{*}/\rm{M_{\odot}}\geq10.5$ (bottom panel) as a function of $D_{\rm fil}$. 
Considering the typical radial scale of cosmic filaments in the local Universe is $2-5\,\rm Mpc$ (\citealt{Galarraga-Espinosa_2020, Aguerri_2026}), we show the radial profiles of the fractions up to $6\,\rm Mpc$ in this figure. 
The numbers at the bottom of each panel are the number of galaxies in each $D_{\rm fil}$ bin with an interval of $1.2\,\rm Mpc$. 
The error bars show the standard binomial error of the fractions. 
We define a reference sample free of the potential influence of filaments by selecting galaxies at $D_{\rm fil}>5\,\rm Mpc$. Their measured fractions and standard error are shown with the dashed line and the grey band. 
The fraction of early-type galaxies increases within $D_{\rm fil}<3.6\,\rm Mpc$ with a clear deviation from the reference sample. 
Similarly, the fraction of massive galaxies is higher close to the filaments.

We quantify the significance of the deviation between the observed fraction and the reference fraction in each bin using Monte Carlo null hypothesis tests, where the null hypothesis is that the observed fraction is drawn from the same underlying distribution as the reference fraction.
For each bin, we first draw a reference fraction from a Gaussian distribution truncated to the interval $[0,1]$, centred on the observed reference fraction and with a standard deviation equal to its binomial standard error, in order to marginalise over the uncertainty in the reference fraction. 
We then generate a mock sample from a binomial distribution with the number of trials set to the number of galaxies in the bin and the success probability set to the sampled reference fraction. 
Repeating this procedure $10^{6}$ times yields a Monte Carlo distribution of mock fractions, from which we estimate the p-value as the probability of obtaining a mock fraction that deviates from the reference fraction by at least as much as the observed fraction. 
Table \ref{tab:early_p_val} shows the p-values in each $D_{\rm fil}$ bin and cumulative intervals. The cumulative p-values are computed using all galaxies within the limits of $D_{\rm fil}$. 
In both early-type and massive galaxy fraction profiles, the p-values are notably higher in bins beyond $D_{\rm fil}\geq3.6\,\rm Mpc$. The cumulative p-values in the $D_{\rm fil}<3.6\,\rm Mpc$ range are 0.0162 and 0.0008 for the early-type and massive galaxy fractions, respectively. This indicates that galaxies close to filaments are significantly more likely to be early-type and massive than galaxies away from filaments ($D_{\rm fil}>5\,\rm Mpc$).

\subsubsection{Filament effects on H{\sc i} detection in late-type galaxies}

\begin{table}
\caption{The same format as Table \ref{tab:early_p_val}, quantifying the significance of the deviation between the observed and reference H{\sc i} detection fractions at different stellar mass and morphology types, shown in Figs. \ref{fig:hi_frac_latetype} and \ref{fig:hi_frac_earlytype}. Only the low-density samples are presented.}
\label{tab:p_val_hi}
\centering
\begin{tabular}{cccc}
\hline
\multicolumn{4}{c}{\textbf{f (H{\sc i} detections) vs $D_{\rm fil}$}} \\ \hline
\multicolumn{4}{c}{\textbf{$8.5 < \log M_{*}/\rm{M_{\odot}} < 9$, Late-type, Low density}} \\ \hline
\multicolumn{2}{c}{Interval statistics} & \multicolumn{2}{c}{Cumulative statistics} \\ \hline
$D_{\rm fil}$ {[}Mpc{]} & p-value & $D_{\rm fil}$ {[}Mpc{]} & p-value \\ \hline
{[}0.0, 1.2) & 0.4289 & - & - \\ \hline
{[}1.2, 2.4) & 0.1727 & {[}0.0, 2.4) & 0.1236 \\ \hline
{[}2.4, 3.6) & 0.0559 & {[}0.0, 3.6) & 0.0300 \\ \hline
{[}3.6, 4.8) & 0.5066 & {[}0.0, 4.8) & 0.0591 \\ \hline
{[}4.8, 6.0) & 0.7937 & {[}0.0, 6.0) & 0.0950 \\ \hline \hline
\multicolumn{4}{c}{\textbf{$9\leq \log M_{*}/\rm{M_{\odot}} < 10$, Late-type, Low density}} \\ \hline
\multicolumn{2}{c}{Interval statistics} & \multicolumn{2}{c}{Cumulative statistics} \\ \hline
$D_{\rm fil}$ {[}Mpc{]} & p-value & $D_{\rm fil}$ {[}Mpc{]} & p-value \\ \hline
{[}0.0, 1.2) & 0.0791 & - & - \\ \hline
{[}1.2, 2.4) & 0.2719 & {[}0.0, 2.4) & 0.0949 \\ \hline
{[}2.4, 3.6) & 0.8431 & {[}0.0, 3.6) & 0.3690 \\ \hline
{[}3.6, 4.8) & 0.2884 & {[}0.0, 4.8) & 0.2857 \\ \hline
{[}4.8, 6.0) & 0.7780 & {[}0.0, 6.0) & 0.3791 \\ \hline \hline
\multicolumn{4}{c}{\textbf{$10\leq \log M_{*}/\rm{M_{\odot}}$, Late-type, Low density}} \\ \hline
\multicolumn{2}{c}{Interval statistics} & \multicolumn{2}{c}{Cumulative statistics} \\ \hline
$D_{\rm fil}$ {[}Mpc{]} & p-value & $D_{\rm fil}$ {[}Mpc{]} & p-value \\ \hline
{[}0.0, 1.2) & - & - & - \\ \hline
{[}1.2, 2.4) & 0.0368 & - & - \\ \hline
{[}2.4, 3.6) & 0.2012 & {[}0.0, 3.6) & 0.0403 \\ \hline
{[}3.6, 4.8) & 0.6577 & {[}0.0, 4.8) & 0.0582 \\ \hline
{[}4.8, 6.0) & 0.7059 & {[}0.0, 6.0) & 0.1256 \\ \hline \hline
\multicolumn{4}{c}{\textbf{$10\leq \log M_{*}/\rm{M_{\odot}}$, Early-type, Low density}} \\ \hline
\multicolumn{2}{c}{Interval statistics} & \multicolumn{2}{c}{Cumulative statistics} \\ \hline
$D_{\rm fil}$ {[}Mpc{]} & p-value & $D_{\rm fil}$ {[}Mpc{]} & p-value \\ \hline
{[}0.0, 1.2) & 0.0627 & - & - \\ \hline
{[}1.2, 2.4) & 0.0267 & {[}0.0, 2.4) & 0.0119 \\ \hline
{[}2.4, 3.6) & 0.4084 & {[}0.0, 3.6) & 0.0349 \\ \hline
{[}3.6, 4.8) & 0.4371 & {[}0.0, 4.8) & 0.0375 \\ \hline
{[}4.8, 6.0) & 1.0000 & {[}0.0, 6.0) & 0.0467 \\\hline
\end{tabular}%
\end{table}

Previously in Section \ref{sec:intrinsic}, we showed that the H{\sc i} detection fraction depends on stellar mass and morphology. To isolate the effect of filaments on H{\sc i} detection from these intrinsic effects, we control for stellar mass and morphology in our analysis. 
In addition, although the local galaxy number density generally increases from voids toward filaments, filament environments span a wide range of local densities, as shown in Fig. \ref{fig:envs}. Since the H{\sc i} detection fraction also depends on local density (Section \ref{sec:local_den_effects}), it is necessary to account for different local environments when assessing the influence of filaments on H{\sc i} detection.

In Fig. \ref{fig:hi_frac_latetype}, we show the H{\sc i} detection fraction of late-type galaxies as a function of $D_{\rm fil}$. The solid lines show the H{\sc i} detection fraction of late-type galaxies in low-local density (upper panels; $\log \Sigma_{5}/\rm{Mpc^{-2}}<0$) and high-local density (middle panels; $\log \Sigma_{5}/\rm{Mpc^{-2}}\geq0$) environments. 
Following \citet{Cameron_2011}, uncertainties on the H{\sc i} detection fraction (error bars associated with the solid lines) are estimated using Jeffrey's interval, a Bayesian credible interval appropriate for binomial statistics with small sample sizes.

The numbers presented at the bottom of each panel are the number of optical galaxies in each bin. 
The dotted line is connected to the data point with low statistical significance, with only one galaxy in the $D_{\rm fil}$ bin. 
The three columns correspond to the stellar mass ranges of $8.5<\log M_{*}/\rm{M_{\odot}}<9$, $9\leq\log M_{*}/\rm{M_{\odot}}<10$, and $\log M_{*}/\rm{M_{\odot}}\geq10$ from left to right. Similarly to Fig. \ref{fig:early_frac}, the dashed horizontal lines and shaded bands show the H{\sc i} detection fraction of reference samples away from the filaments ($D_{\rm fil}>5\,\rm Mpc$), with the same mass and local density conditions. 

Dividing the galaxies into three stellar mass bins partially controls the stellar mass effects. To test that stellar mass variations within each bin do not drive the observed trends with $D_{\rm fil}$, we show the mean and 1$\sigma$ scatter of the stellar mass of the low- and high-local density galaxies in the bottom panels. 
We do not find a significant variation in the overall distribution of stellar mass as a function of $D_{\rm fil}$ that might strongly affect the H{\sc i} fraction seen in the upper panels. Potential exceptions are the inner-most bin of the low-mass low-density sample, where the stellar mass of the only galaxy in this bin is lower than the mean stellar mass of galaxies in other bins, and the inner-most bin of the high-mass high-density sample, where the sample mean is slightly higher than the other bins.

In the lowest stellar mass bin, $8.5<\log M_{*}/\rm{M_{\odot}}<9$ (top left panel), we find reduced H{\sc i} detection fractions across three inner bins. 
This is supported by the p-value measurements presented in Table \ref{tab:p_val_hi}, such that the p-values of the null hypothesis in these bins are smaller than those in other bins at larger $D_{\rm fil}$. 
The statistical significance of the deviation across the three bins has a cumulative p-value of 0.03. 
The level of H{\sc i} detection fraction within $D_{\rm fil}<3.6\,\rm Mpc$, $34^{+9}_{-8}\%$, is even slightly lower than the high-local density sample in the same stellar mass and $D_{\rm fil}$ ranges ($45\pm 6\%$; centre left panel). 
Earlier in Fig. \ref{fig:detection_localden}, we show that the H{\sc i} detection fraction of late-type galaxies depends on local density, such that galaxies in higher local densities are less likely to have H{\sc i} detection. While this is to some degree reflected in the H{\sc i} detection fractions of the reference samples ($57\pm 7\%$ and $43\pm10\%$ for the low- and high-density samples, respectively), the lower H{\sc i} detection fraction among low-local density galaxies in filaments indicates that the local density dependency of the H{\sc i} detection fraction vanishes close to filaments.

In the low-density intermediate mass sample ($9\leq\log M_{*}/\rm{M_{\odot}}<10$; top centre panel), we find the H{\sc i} detection fraction of the innermost bin ($25^{+18}_{-12}\%$) is smaller than that of the reference sample ($57\pm6\%$). 
Compared to the lower stellar mass galaxies in the left panel, the reduced H{\sc i} detection is only visible at a smaller distance from the filaments, within $D_{\rm fil}<1.2\,\rm Mpc$ (see also p-values in Table \ref{tab:p_val_hi}). 
The p-value for this deviation is 0.0791, indicating only tentative evidence for a decreased H{\sc i} detection near filaments from the small number statistic (8 galaxies).  

In the highest stellar mass bin ($\log M_{*}/\rm{M_{\odot}}\geq10$), there are no low-local density late-type galaxies within $D_{\rm fil}<1.2\,\rm Mpc$. Yet, we still find reduced H{\sc i} detection fractions in $1.2\leq D_{\rm fil}/\rm Mpc<2.4$ ($42^{+14}_{-13}\%$) and potentially $2.4\leq D_{\rm fil}/\rm Mpc<3.6$ ($58^{+13}_{-14}\%$) ranges, compared to that of the reference sample ($75\pm9\%$).
The cumulative p-value of the deviation across these two bins is 0.0403. 

We do not find any $D_{\rm fil}$ dependency among the high-local density samples presented in panels in the centre row. 
Regardless of the stellar mass range, the H{\sc i} detection fractions of these galaxies remain nearly constant with varying $D_{\rm fil}$ and are statistically indistinguishable from the reference samples away from the filaments.

\subsubsection{Filament effects on H{\sc i} detection in early-type galaxies}

Fig. \ref{fig:hi_frac_earlytype} shows the H{\sc i} detection fraction of early-type galaxies, in the same format as Fig. \ref{fig:hi_frac_latetype}. 
We only show the highest stellar mass bin ($\log M_{*}/\rm{M_{\odot}}\geq10$), as there are few early-type galaxies at the smaller mass ranges. 
A higher local density limit is used to separate low-density ($\log \Sigma_{5}/\rm{Mpc^{-2}}<0.5$) and high-density ($\log \Sigma_{5}/\rm{Mpc^{-2}}\geq0.5$) samples as early-type galaxies generally occupy higher local density environments compared to late-type galaxies (see Fig. \ref{fig:envs}). 
While most early-type galaxies in our sample are not detected in H{\sc i}, we find an interesting behaviour among massive early-type galaxies in low-density environments close to filaments (blue line with error bars in the top panel). 
The H{\sc i} detection fraction of this sample tends to increase with decreasing filament distance, from $14\pm 9\%$ in $D_{\rm fil}>5\,\rm Mpc$ to $39\pm10\%$ in $1.2\leq D_{\rm fil}<2.4$ and $50\pm 19\%$ in the innermost bin of $D_{\rm fil}<1.2\,\rm Mpc$. 
The large error margin is due to the limited number of early-type galaxies in our sample. Yet, the cumulative p-value of the deviation in $D_{\rm fil}<2.4\,\rm Mpc$ compared to the reference sample is 0.0119, suggesting that the observed trend is statistically robust. 
As shown in the bottom panel, the average stellar mass of galaxies remains nearly constant as a function of $D_{\rm fil}$, thus, it is unlikely that the increased H{\sc i} detection fraction is caused by the H{\sc i} mass--stellar mass relation. 
We do not find such deviations from the reference sample among galaxies in high-local densities (middle panel).

\section{Discussion}\label{sec:discussion}

\subsection{Filaments as a driver of galaxy evolution}

As indicated by the wide range of local densities occupied by galaxies in filaments (see Fig. \ref{fig:envs}), galaxies are not uniformly distributed along filaments but arranged in alternating high- and low-density regions (\citealt{Tempel_2014b}). 
Theoretical studies demonstrate that
star formation, gas content, and the kinematics of galaxies are related to the matter density of the nearest filament segment (\citealt{Hasan_2024, Hasan_2025}).
Filament proximity and local galaxy number density should therefore be considered as related but distinct environmental measures. The former traces the location of galaxies within the LSS network, while the latter quantifies their immediate surrounding environments. 

In this study, we show that the H{\sc i} detection fraction of galaxies depends on stellar mass, morphology, and local density, all of which vary with filament proximity. Galaxies closer to filaments tend to have higher stellar masses, reside in higher local densities, and show a larger fraction of early-type morphologies. 
Similar dependencies of galaxy properties on filament environments have been reported in previous studies, although it has been regarded that some of these dependencies, if not all, may be driven by the higher local densities of filament regions compared to voids (e.g., \citealt{Kraljic_2018, OKane_2024, Barsanti_2025, Gallagher_2026}). 

However, our results show that filament proximity still has a measurable effect on galaxies, particularly on H{\sc i} detection, even after controlling for intrinsic galaxy properties and local density.  
In the following sections, we interpret our observations within this framework: local density regulates galaxy–galaxy and galaxy–halo interactions between immediate neighbours, while filament proximity is linked to gas supply or removal mechanisms within the cosmic web.

\subsection{Physical interpretation}

The observed variations in the H{\sc i} detection fractions as a function of the distance to filaments are an outcome of several competing processes operating in and around filaments that lead galaxy evolution. H{\sc i} is a particularly sensitive tracer of these processes because changes in H{\sc i} gas often occur in earlier stages of transformation, compared to morphology or star formation rate, which may respond only after substantial changes in gas properties have already taken place. For example, gas stripping, starvation, or suppressed gas accretion can leave an imprint on H{\sc i} properties even when galaxies retain late-type morphologies or sustain their star formation rate (\citealt{Chung_2009, Brown_2017}).

\subsubsection{Gas supply mechanisms}

The redshift range investigated in this study ($0.02<z<0.09$) is often considered to be dominated by hot-mode accretion of gas (e.g., \citealt{Keres_2005, Correa_2018, Wright_2021}), but in fact, cold-mode filamentary accretion can still take place in low-density void environments. 
Whether a halo of a certain mass can be a dominant hub for attracting matter in the cosmic web can be approximated by the characteristic non-linear mass at a given redshift and environment. 
\citet{Aragon-Calvo_2019} show that the non-linear mass of a halo given the mean density of the Universe at $z=0$ is on the scale of $\sim 10^{13}\, \mathrm{M_{\odot}}\,h^{-1}$, which typically hosts $\sim 10^{11}\, \mathrm{M_{\odot}}h^{-1}$ stellar mass galaxies. Meanwhile, in the void regions where the density is lower than the universal average, this decreases to $\lesssim 10^{12}\, \mathrm{M_{\odot}}\,h^{-1}$ (i.e., $\lesssim 10^{10}\, \mathrm{M_{\odot}}\,h^{-1}$ in stellar mass). 
The non-linear mass also decreases with increasing redshift, meaning even smaller galaxies can be a node of local matter flow in low-density environments throughout the evolution. 
Considering the mass range of our galaxy sample, we conclude that the high H{\sc i} detection fraction we observe among late-type galaxies in low-local density environments can be an outcome of past and ongoing gas accretion from tenuous tendril-like filaments onto individual galaxies. 
Such low-density structures of the cosmic web are not captured by our filament-finding procedure, which is designed to identify high-significance filaments traced by multiple galaxies with stellar mass larger than $\log M_{*}/\rm{M_{\odot}}>9$ (see Section \ref{sec:filament_finding}). 
The gas accretion scenario is supported by the high H{\sc i} content and the consequent H{\sc i} depletion time of MIGHTEE-H{\sc i} Early Science Release galaxies presented in \citet{Tudorache_2026}.


Specific to cosmic filaments, \citet{Kleiner_2017} report that massive galaxies ($\log M_{*}/\rm{M_{\odot}}>11$) at $z < 0.043$ located at the core of cosmic filaments have higher H{\sc i} mass fractions than the control sample and attribute this result to the cold-mode accretion of gas from filaments. 
Similarly, we find an increase in the H{\sc i} detection fraction among massive galaxies with early-type morphology close to low-local density sections of filaments. However, this trend is not visible in higher local densities, and massive late-type galaxies instead show reduced H{\sc i} detections close to filaments. 
Our results suggest that either (i) early-type galaxies are more efficient at receiving gas from filaments than late-type galaxies, or (ii) there are other gas supply mechanisms active in early-type galaxies, such as gas-rich mergers (\citealt{Sancisi_2008, Serra_2012, Tudorache2022}), which we discuss below.

\subsubsection{Mergers}
We find an increased fraction of early-type galaxies within $<3.6\,\rm Mpc$ of filaments. This is consistent with previous studies showing that mergers and strong gravitational interactions are frequent in filamentary environments (e.g. \citealt{Kuutma_2017, Jung_2025, Dulcien_2026}), where galaxies are channelled from voids and sheets and move toward denser structures. Such interactions can drive the morphological transformation from late-type to early-type. 

Away from filaments, early-type galaxies generally have a lower H{\sc i} detection fraction than late-type galaxies. This indicates that morphological transformation is often accompanied by the depletion of H{\sc i}, and that galaxies rarely replenish their cold gas content to levels detectable in our observations. 
Interestingly, we find a different behaviour among massive galaxies in the low-density parts of filaments; early-type galaxies have H{\sc i} detection fractions as high as their late-type counterparts, suggesting potential gas replenishment in this environment.

\subsubsection{Ram pressure stripping}

One possible mechanism for gas removal from low-mass galaxies in filaments is ram pressure stripping (\citealt{Benitez-Llambay_2013, Herzog_2023}). 
Since ram pressure stripping is a hydrodynamic process, it can affect the galactic gas without morphological transformation. 
Comparing late-type galaxies in stellar mass ranges of $8.5<\log M_{*}/\rm{M_{\odot}}<9$ and $9\leq\log M_{*}/\rm{M_{\odot}}<10$, we find that both populations show reduced H{\sc i} detection fractions close to filaments. However, the decline begins at larger filament distances for the lower-mass galaxies ($<3.6\,\rm Mpc$) than for the higher-mass galaxies ($<1.2\,\rm Mpc$). 
This is consistent with the expectation that the effective radius of ram pressure stripping depends on galaxy mass, as the efficiency of gas stripping is governed by the balance between ram pressure and the gravitational restoring force.


The decline in H{\sc i} detection fraction is found primarily among galaxies in low-local-density regions in filaments. These galaxies are potentially infalling into filaments from voids or sheets as isolated individual galaxies, making them more directly exposed to the filament medium.  
In contrast, we do not find strong evidence for ram pressure stripping among galaxies in high-density environments. 
If these galaxies interact with filaments as members of gravitationally bound groups, ram pressure stripping by the filament medium may be less efficient because (i) the deeper gravitational potential of the host group halo can help retain gas within the group, and (ii) the intra-group medium can shield galaxies from direct hydrodynamic interaction with the filament gas by reducing their relative velocity with respect to the surrounding medium. 
As a result, their cold gas reservoirs may be less susceptible to direct stripping by the filaments.

We note that even galaxies in the highest stellar mass bin ($\log M_{*}/\rm{M_{\odot}}>10$) show a significant decrease in H{\sc i} detection fraction close to filaments. However, this trend is unlikely to be explained solely by ram pressure stripping, considering that massive galaxies have stronger gravitational restoring forces.
Other processes may contribute to the reduced H{\sc i} detection of massive galaxies near filaments. We will discuss this in more detail in the next section.

As mentioned earlier, we explicitly exclude galaxies within $\leq2\,\rm Mpc$ from nodes in our analysis presented in Section \ref{sec:fil_effects} to separate filament-related environmental effects from those associated with nodes. Nevertheless, we note that some galaxies in our sample are located on the outskirts of nodes where massive groups reside. 
In such environments, ram pressure stripping can already be in effect by the extended intracluster medium beyond several virial radii (e.g., \citealt{Bahe_2013}). 
Nevertheless, we confirm that applying a stricter node-distance cut (e.g., $D_{\rm node}>5\,\rm Mpc$) does not qualitatively change the dependence of the H{\sc i} detection fraction on filament proximity. This suggests that the trends discussed in this paper are not driven solely by galaxies associated with nodes.

\subsubsection{Cosmic web detachment}

In addition to direct gas removal from galaxies, our results are consistent with the cosmic web detachment model (\citealt{Aragon-Calvo_2019}). 
When galaxies in high-density environments experience interactions with neighbouring galaxies or halos, it can lead to detachment from the primordial tendrils attached to these galaxies and, as a result, to a gradual depletion of gas reservoirs. 
This may explain the lower H{\sc i} detection fractions observed among late-type galaxies in high-local-density environments across all stellar mass bins, compared to galaxies residing in lower-density regions.

Since galaxies that formed in these high-density environments are likely to have already experienced detachment from a primordial filamentary gas supply, subsequent interactions with filaments at later stages of their evolution may have a limited impact on their gas accretion history. This explains our observation that the H{\sc i} detection fraction at high-local densities does not depend on the filament proximity. 
The presence of detectable H{\sc i} in these systems is rather likely to depend on a combination of stochastic gas accretion events and internal processes, including feedback-driven heating or gas cooling from the hot halo.

In contrast, when galaxies in low-density environments that are attached with gas tendrils start to interact with a large-scale structure such as filaments, they will experience gas depletion following the cosmic web detachment and/or inefficient accretion of gas onto galaxies (\citealt{Song_2021}).
Since these processes operate over long timescales, their effects can be more apparent among massive galaxies which are likely to have experienced shell-crossing events earlier in the assembly of the cosmic web and had sufficient time for gas depletion. 
This could explain the decline in the H{\sc i} detection fraction observed among massive low-density late-type galaxies near filaments.

\subsection{Extracting information from H{\sc i} non-detections}

In this study, we use the H{\sc i} detection fraction as a measure of the galactic cold gas reservoir and show that it varies with various intrinsic galaxy properties as well as cosmic environments. 
Given the observational sensitivity of the MIGHTEE-H{\sc i} survey and source-finding criteria, it is important to point out that an H{\sc i} non-detection of an optical galaxy does not necessarily mean the complete absence of H{\sc i} in the galaxy. Instead, any galaxies with reduced H{\sc i} emission line flux due to low H{\sc i} mass, large distance, large inclination, and wide velocity width can belong to this category (\citealt{Maksymowicz-Maciata_2026}). 
Therefore, the H{\sc i} detection fraction does not fully capture the changes in the underlying distribution of H{\sc i} masses. 

This limitation is closely related to the motivation for H{\sc i} stacking (e.g., \citealt{Sinigaglia_2024}). 
In \citet{Pan_2025}, the authors reconstruct the distribution of H{\sc i} mass at a given stellar mass using H{\sc i} flux measurements extracted at the positions of optically selected galaxies, regardless of whether each galaxy has statistically significant H{\sc i} emission above the observational noise level. 
In this way, detections and non-detections are treated equally, and the resulting H{\sc i} mass distributions are significantly extended towards low H{\sc i} masses. In other words, these low H{\sc i} mass galaxies are non-detections when considered individually, but provide valuable information about the low mass tail of the H{\sc i} mass distribution when stacked together. 

The fraction of H{\sc i} detections and its dependence on the environments presented in this study hint that the underlying H{\sc i} mass distribution, and therefore the stellar mass -- H{\sc i} mass relation, varies with environment. 
Stacking of H{\sc i} flux across various cosmic environments is beyond the scope of this study, but it is an interesting direction for future studies. 
It could provide insight into gas depletion mechanisms in filaments. 
For example, a lower H{\sc i} detection fraction within filaments could be an outcome of (i) an overall downward shift of the H{\sc i} masses and/or (ii) a broader distribution with a low-H{\sc i} mass tail. 
The former can be interpreted as a gradual gas loss over an extended time, while the latter can be a result of a more rapid, stochastic removal of gas. 

Furthermore, investigating the population of star-forming galaxies with non-H{\sc i} detections could offer an opportunity to study the timescale of star formation quenching with respect to the timescale of gas depletion. If these galaxies are genuinely H{\sc i}-deficient rather than non-detections due to the survey sensitivity, they represent systems observed in a transitional phase between H{\sc i} gas depletion and quenching. Their abundance provides a constraint on the timescale of environmentally driven quenching.

\section{Summary}\label{sec:summary}

In this paper, we analyse the distribution of H{\sc i} detections with respect to optical galaxy populations. 
We use the DESI DR1 stellar mass and emission line catalogue to construct a sample of 865 optical galaxies. Their H{\sc i} detections are evaluated using a catalogue from the MIGHTEE-H{\sc i} survey in the COSMOS and XMM-LSS fields. The cosmic filament identification is performed by applying the DisPerSE algorithm to the DESI spectroscopic galaxy distribution. 
The mass-limited galaxy samples span a wide range of stellar masses ($8.5<\log M_{*}/\rm{M_{\odot}}<11.3$) and local densities ($-1<\log\Sigma_{\rm 5}/\rm{Mpc^{-2}}<3.1$), and diverse cosmic web environments (groups, filaments, and voids). 

It is important to note that the H{\sc i} sources are identified directly from the radio spectral data cubes using untargeted source finding, without prior selection based on the optical galaxy distribution. The H{\sc i} detections, therefore, provide an independent probe of H{\sc i}-rich galaxies within the MIGHTEE-H{\sc i} survey footprint investigated in this study. 
As such, we use the fraction of H{\sc i} detections in an optically selected sample to quantify the H{\sc i} characteristic. 

Our main findings are summarised as follows. 
First, the H{\sc i} detection fraction depends on intrinsic galaxy properties, such as stellar mass and morphology. The H{\sc i} detection fraction of massive galaxies  ($\log M_{*}/\rm{M_{\odot}}>10$) is particularly low, due to the low fraction of star-forming galaxies in this stellar mass range. 
Early-type galaxies have lower H{\sc i} detection fractions than late-type galaxies at all stellar mass ranges probed in this study. 
Motivated by our finding that the fraction of early-type and massive galaxies increases toward filaments, we control for these intrinsic galaxy properties when assessing the impact of filaments on the H{\sc i} detection.

Second, the local galaxy number density and the cosmic filament proximity are related, but distinct environmental measures that trace different scales. 
While the average local density increases closer to filaments, galaxies near filaments still inhabit a wide range of local densities, reflecting the non-uniform distribution of galaxies along filaments. We find that the H{\sc i} detection fraction depends on local density, even when controlling for galaxy morphology.

Finally, we show that cosmic filaments influence the H{\sc i} detection fraction through various, competing processes whose relative importance depends on stellar mass, morphology, and local environment. Late-type galaxies generally show reduced H{\sc i} detection fractions close to filaments, particularly in low-local density environments. For lower-mass galaxies ($\log M_{*}/\rm{M_{\odot}}<10$), this trend can be explained with gas removal by ram pressure stripping, while for massive galaxies ($\log M_{*}/\rm{M_{\odot}}\geq10$) it may reflect slow gas depletion following cosmic web detachment. In contrast, massive early-type galaxies in low-local density sections of filaments show enhanced H{\sc i} detection fractions, potentially indicating gas replenishment through gas-rich mergers. 

Our results suggest that filaments do not exert a single unified effect on galaxies. Instead, their impact depends on various physical factors such as stellar mass, local density, and whether galaxies interact with filaments as isolated systems or as members of larger groups.

\section*{Acknowledgements}
SLJ, MJJ, TY and BY acknowledge the support of a UKRI Frontiers Research Grant [EP/X026639/1], which was selected by the European Research Council, and the STFC consolidated grants [ST/S000488/1] and [ST/W000903/1]. 
MB acknowledges the financial support from the Flemish Fund for Scientific Research (FWO-Vlaanderen) and the South African National Research Foundation (NRF) under their Bilateral Scientific Cooperation program (grant G0G0420N).
MG is supported through UK STFC Grant ST/Y001117/1. MG acknowledges support from the Inter-University Institute for Data Intensive Astronomy (IDIA). IDIA is a partnership of the University of Cape Town, the University of Pretoria and the University of the Western Cape. For the purpose of open access, the author has applied a Creative Commons Attribution (CC BY) licence to any Author Accepted Manuscript version arising from this submission.
KS acknowledges support from the Natural Sciences and Engineering Research Council of Canada (NSERC) and the Canada Research Chairs program.

Our analysis was performed using the Python programming language (Python Software Foundation, https://www.python.org). The following packages were used throughout the analysis: numpy (\citealt{Harris_2020}), SciPy (\citealt{Virtanen_2020}), matplotlib (\citealt{Hunter_2007}). This work also made use of Astropy:\footnote{http://www.astropy.org} a community-developed core Python package and an ecosystem of tools and resources for astronomy \citep{astropy:2013, astropy:2018, astropy:2022}.

The MeerKAT telescope is operated by the South African Radio Astronomy Observatory, which is a facility of the National Research Foundation, an agency of the Department of Science and Innovation. We acknowledge the use of the ilifu cloud computing facility – www.ilifu.ac.za, a partnership between the University of Cape Town, the University of the Western Cape, Stellenbosch University, Sol Plaatje University and the Cape Peninsula University of Technology. The Ilifu facility is supported by contributions from the Inter-University Institute for Data Intensive Astronomy (IDIA – a partnership between the University of Cape Town, the University of Pretoria and the University of the Western Cape, the Computational Biology division at UCT and the Data Intensive Research Initiative of South Africa (DIRISA). The authors acknowledge the Centre for High Performance Computing (CHPC), South Africa, for providing computational resources to this research project.

This research used data obtained with the Dark Energy Spectroscopic Instrument (DESI). DESI construction and operations is managed by the Lawrence Berkeley National Laboratory. This material is based upon work supported by the U.S. Department of Energy, Office of Science, Office of High-Energy Physics, under Contract No. DE–AC02–05CH11231, and by the National Energy Research Scientific Computing Center, a DOE Office of Science User Facility under the same contract. Additional support for DESI was provided by the U.S. National Science Foundation (NSF), Division of Astronomical Sciences under Contract No. AST-0950945 to the NSF’s National Optical-Infrared Astronomy Research Laboratory; the Science and Technology Facilities Council of the United Kingdom; the Gordon and Betty Moore Foundation; the Heising-Simons Foundation; the French Alternative Energies and Atomic Energy Commission (CEA); the National Council of Humanities, Science and Technology of Mexico (CONAHCYT); the Ministry of Science and Innovation of Spain (MICINN), and by the DESI Member Institutions: www.desi.lbl.gov/collaborating-institutions. The DESI collaboration is honored to be permitted to conduct scientific research on I’oligam Du’ag (Kitt Peak), a mountain with particular significance to the Tohono O’odham Nation. Any opinions, findings, and conclusions or recommendations expressed in this material are those of the author(s) and do not necessarily reflect the views of the U.S. National Science Foundation, the U.S. Department of Energy, or any of the listed funding agencies.

The Legacy Surveys consist of three individual and complementary projects: the Dark Energy Camera Legacy Survey (DECaLS; Proposal ID \#2014B-0404; PIs: David Schlegel and Arjun Dey), the Beijing-Arizona Sky Survey (BASS; NOAO Prop. ID \#2015A-0801; PIs: Zhou Xu and Xiaohui Fan), and the Mayall z-band Legacy Survey (MzLS; Prop. ID \#2016A-0453; PI: Arjun Dey). DECaLS, BASS and MzLS together include data obtained, respectively, at the Blanco telescope, Cerro Tololo Inter-American Observatory, NSF’s NOIRLab; the Bok telescope, Steward Observatory, University of Arizona; and the Mayall telescope, Kitt Peak National Observatory, NOIRLab. Pipeline processing and analyses of the data were supported by NOIRLab and the Lawrence Berkeley National Laboratory (LBNL). The Legacy Surveys project is honored to be permitted to conduct astronomical research on Iolkam Du’ag (Kitt Peak), a mountain with particular significance to the Tohono O’odham Nation.

NOIRLab is operated by the Association of Universities for Research in Astronomy (AURA) under a cooperative agreement with the National Science Foundation. LBNL is managed by the Regents of the University of California under contract to the U.S. Department of Energy.

This project used data obtained with the Dark Energy Camera (DECam), which was constructed by the Dark Energy Survey (DES) collaboration. Funding for the DES Projects has been provided by the U.S. Department of Energy, the U.S. National Science Foundation, the Ministry of Science and Education of Spain, the Science and Technology Facilities Council of the United Kingdom, the Higher Education Funding Council for England, the National Center for Supercomputing Applications at the University of Illinois at Urbana-Champaign, the Kavli Institute of Cosmological Physics at the University of Chicago, Center for Cosmology and Astro-Particle Physics at the Ohio State University, the Mitchell Institute for Fundamental Physics and Astronomy at Texas A\&M University, Financiadora de Estudos e Projetos, Fundacao Carlos Chagas Filho de Amparo, Financiadora de Estudos e Projetos, Fundacao Carlos Chagas Filho de Amparo a Pesquisa do Estado do Rio de Janeiro, Conselho Nacional de Desenvolvimento Cientifico e Tecnologico and the Ministerio da Ciencia, Tecnologia e Inovacao, the Deutsche Forschungsgemeinschaft and the Collaborating Institutions in the Dark Energy Survey. The Collaborating Institutions are Argonne National Laboratory, the University of California at Santa Cruz, the University of Cambridge, Centro de Investigaciones Energeticas, Medioambientales y Tecnologicas-Madrid, the University of Chicago, University College London, the DES-Brazil Consortium, the University of Edinburgh, the Eidgenossische Technische Hochschule (ETH) Zurich, Fermi National Accelerator Laboratory, the University of Illinois at Urbana-Champaign, the Institut de Ciencies de l’Espai (IEEC/CSIC), the Institut de Fisica d’Altes Energies, Lawrence Berkeley National Laboratory, the Ludwig Maximilians Universitat Munchen and the associated Excellence Cluster Universe, the University of Michigan, NSF’s NOIRLab, the University of Nottingham, the Ohio State University, the University of Pennsylvania, the University of Portsmouth, SLAC National Accelerator Laboratory, Stanford University, the University of Sussex, and Texas A\&M University.

BASS is a key project of the Telescope Access Program (TAP), which has been funded by the National Astronomical Observatories of China, the Chinese Academy of Sciences (the Strategic Priority Research Program ``The Emergence of Cosmological Structures'' Grant \# XDB09000000), and the Special Fund for Astronomy from the Ministry of Finance. The BASS is also supported by the External Cooperation Program of Chinese Academy of Sciences (Grant \# 114A11KYSB20160057), and Chinese National Natural Science Foundation (Grant \# 12120101003, \# 11433005).

The Legacy Survey team makes use of data products from the Near-Earth Object Wide-field Infrared Survey Explorer (NEOWISE), which is a project of the Jet Propulsion Laboratory/California Institute of Technology. NEOWISE is funded by the National Aeronautics and Space Administration.

The Legacy Surveys imaging of the DESI footprint is supported by the Director, Office of Science, Office of High Energy Physics of the U.S. Department of Energy under Contract No. DE-AC02-05CH1123, by the National Energy Research Scientific Computing Center, a DOE Office of Science User Facility under the same contract; and by the U.S. National Science Foundation, Division of Astronomical Sciences under Contract No. AST-0950945 to NOAO. 
The Photometric Redshifts for the Legacy Surveys (PRLS) catalog used in this paper was produced thanks to funding from the U.S. Department of Energy Office of Science, Office of High Energy Physics via grant DE-SC0007914.

This research made use of the ``K-corrections calculator'' service available at http://kcor.sai.msu.ru/

\section*{Data availability}
The MIGHTEE-H{\sc i} source catalogue is available as supplementary material of \citet{Maksymowicz-Maciata_2026}. 
The DESI DR1 stellar mass and emission line catalogue is available at \url{https://data.desi.lbl.gov/doc/releases/dr1/vac/stellar-mass-emline/}. The Photometric Redshift files for the Legacy Surveys are available at \url{https://portal.nersc.gov/cfs/cosmo/data/legacysurvey/dr8/south/sweep/8.0-photo-z/}.



\bibliographystyle{mnras}
\bibliography{references} 

\end{document}